\documentclass[sigconf]{acmart}
\usepackage{graphicx}
\usepackage{subfig}
\usepackage{caption}
\usepackage{color}
\usepackage{mathtools}
\usepackage{url}
\usepackage{hyperref}
\usepackage{bm}
\usepackage{dsfont}
\usepackage{multirow}
\usepackage{bbding}

\AtBeginDocument{%
  }

\copyrightyear{2022}
\acmYear{2022}
\setcopyright{acmlicensed}\acmConference[SIGIR '22]{Proceedings of the 45th International ACM SIGIR Conference on Research and Development in Information Retrieval}{July 11--15, 2022}{Madrid, Spain}
\acmBooktitle{Proceedings of the 45th International ACM SIGIR Conference on Research and Development in Information Retrieval (SIGIR '22), July 11--15, 2022, Madrid, Spain}
\acmPrice{15.00}
\acmDOI{10.1145/3477495.3532044}
\acmISBN{978-1-4503-8732-3/22/07}
\begin{document}
\fancyhead{}
\title{Privacy-Preserving Synthetic Data Generation for Recommendation Systems}
\author{Fan Liu}
\affiliation{%
  \institution{National University of Singapore}
  \country{}
  }
\email{liufancs@gmail.com}

\author{Zhiyong Cheng}
\affiliation{%
  \institution{Qilu University of Technology (Shandong Academy of Sciences)}
  \country{}
  }
\email{jason.zy.cheng@gmail.com}

\author{Huilin Chen}
\affiliation{%
  \institution{Tianjin University of Technology}
  \country{}
  }
\email{ClownClumsy@outlook.com}

\author{Yinwei Wei}
\affiliation{%
  \institution{National University of Singapore}
  \country{}
  }
\email{weiyinwei@hotmail.com}

\author{Liqiang Nie}
\affiliation{%
  \institution{Shandong University}
  \country{}
  }
\email{nieliqiang@gmail.com}

\author{Mohan Kankanhalli}
\affiliation{%
  \institution{National University of Singapore}
  \country{}
  }
\email{mohan@comp.nus.edu.sg}


\begin{abstract}
Recommendation systems make predictions chiefly based on users' historical interaction data ($e.g.$, items previously \emph{clicked} or \emph{purchased}).
There is a risk of privacy leakage when collecting the users' behavior data for building the recommendation model. However,
existing privacy-preserving solutions are designed for tackling the privacy issue only during the model training~\cite{Yang2019FRS} and results collection~\cite{Wang2017DPRS} phases. 
The problem of privacy leakage still exists when directly sharing the private user interaction data with organizations or releasing them to the public. 
To address this problem, in this paper, we present a User Privacy Controllable Synthetic Data Generation model (short for UPC-SDG), which generates synthetic interaction data for users based on their privacy preferences.  
The generation model aims to provide certain privacy guarantees while maximizing the utility of the generated synthetic data at both data level and item level. Specifically,
at the data level, we design a selection module that selects those items that contribute less to a user's preferences from the user's interaction data. At the item level, a synthetic data generation module is proposed to generate a synthetic item corresponding to the selected item based on the user's preferences. Furthermore, we also present a privacy-utility trade-off strategy to balance the privacy and utility of the synthetic data. Extensive experiments and ablation studies have been conducted on three publicly accessible datasets to justify our method, demonstrating its effectiveness in generating synthetic data under users' privacy preferences.
\end{abstract}



\maketitle
\section{Introduction}
Recommendation systems, particularly collaborative filtering (CF) systems~\cite{Rendle2009Bpr,He2017Neural,He2020lightgcn,Liu2020A2GCN,liu2021interest,liu2022review}, play an important role in various online platforms (e.g., Amazon\footnote{https://www.amazon.com}, eBay\footnote{https://www.eBay.com}, TikTok\footnote{https://www.tiktok.com\label{tiktok}}). It makes personalized recommendations of products or services to users to help them navigate through an overwhelming amount of information. Most existing CF-based recommendation methods learn the users' and items' representations by exploiting the information from the users' historical interaction data, such as the behavioral information, purchase history, recommendation feedback, etc. 
As the recommendation methods are heavily data-driven, the more data it uses, the better is the obtained recommendation~\cite{He2017Neural}. However, such users' data is often highly private for numerous personal, social, and financial reasons. 
The risks concerning violating individuals' privacy present a significant impediment to storing or using the users' private data in an insecure environment. For instance, the user's private data may be shared with various organizations or stored in the public cloud.

In the past few years, privacy issues in recommendation systems have been attracting more attention. Many emerging solutions have been developed by applying privacy-preserving techniques (e.g., Federated learning~\cite{Yang2019FRS} and Differential privacy~\cite{Wang2017DPRS,Gao2020DPLCF}) in recommendation systems. For example, Yang et al.~\cite{Yang2019FRS} proposed a federated learning-based approach, which collaboratively trains recommendation models with users' private data maintained locally at each party (mobile device or organization). The user's private data is stored locally by each party in their approach. Only the intermediate results (e.g., parameter updates) are used to communicate with other parties. Gao et al.~\cite{Gao2020DPLCF} adopted a differentially private protection mechanism to help obfuscate the real interactions before reporting them to the server in the training process. Wang et al.~\cite{Wang2017DPRS} applied the differential privacy concept to prevent user history information recovery through the recommendation results. In general, the existing solutions are presented based on the decentralized framework. Besides, they are designed for the privacy issue in the model training~\cite{Yang2019FRS,Gao2020DPLCF} and result collection~\cite{Wang2017DPRS} phases.

Despite their effectiveness, we argue that the existing solutions still suffer from the following three limitations: (1) \textbf{Communication and computation cost}. Data transfers and local computing in decentralized methods make it hard to apply these methods in real-world recommendation scenarios; (2) \textbf{Data sharing or releasing risks}. There is a considerable risk of privacy leakage while explicitly sharing data with other organizations or storing the raw data in public (e.g., the public cloud); (3) \textbf{Users have different privacy preferences for different recommendation scenarios}. For example, users would pay more attention to their medical and financial information privacy protection than for grocery items purchase records. Existing solutions neglect users' privacy preferences while building recommendation system.
\begin{figure}[]
		\centering 
		\includegraphics[width=0.9\linewidth]{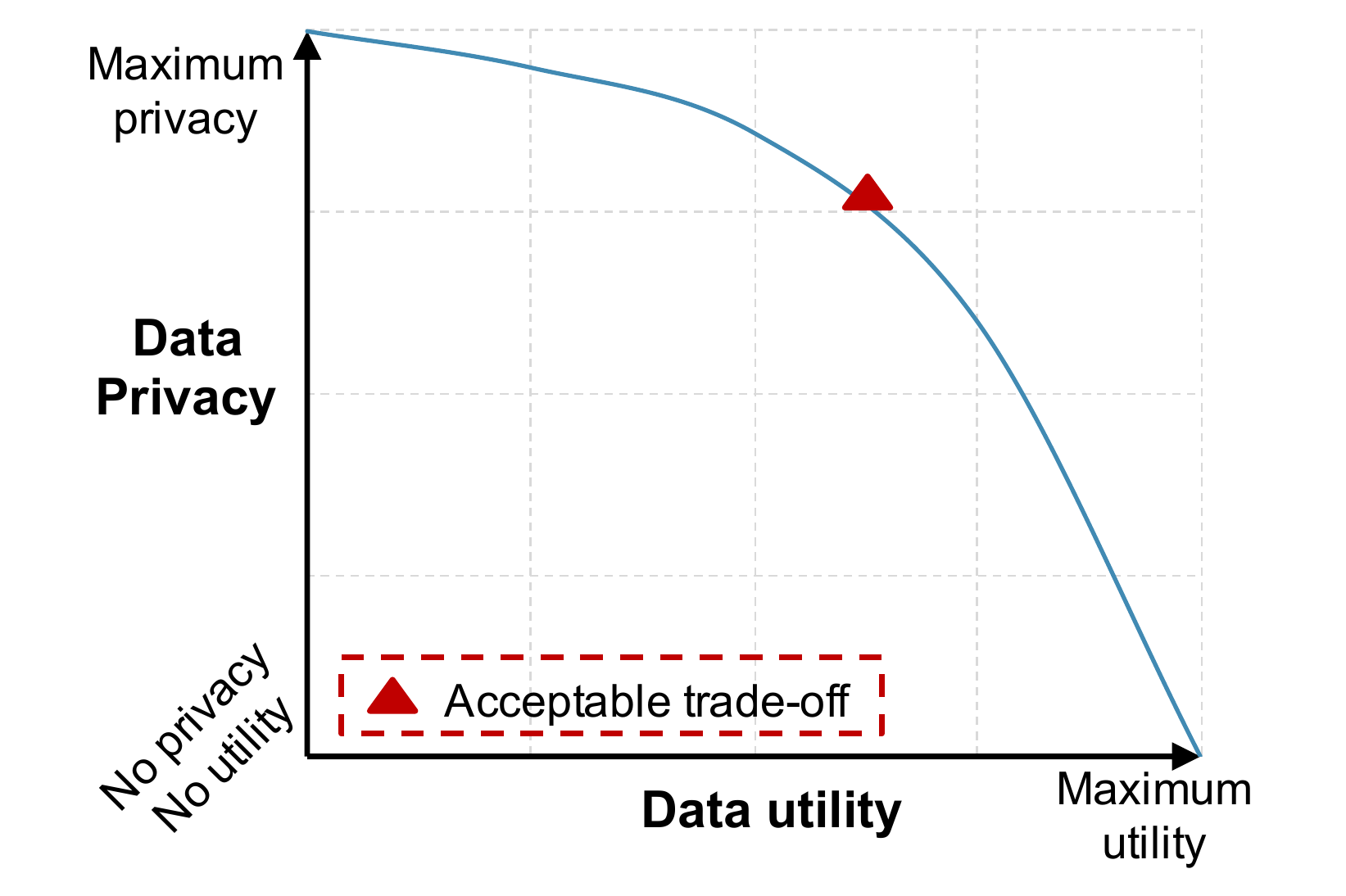} 
		\caption{The trade-off between data privacy and data utility.}  
		\label{fig:trade_off}  
		\vspace{0.0cm}
\end{figure}

The use of privacy-preserving synthetic data provides an alternative method to tackle the privacy issue in the recommendation. The key idea of privacy-preserving synthetic data is to create synthetic data from generative models while preserving as many properties of the original data as possible. As shown in Fig.~\ref{fig:trade_off}, there is a trade-off between data privacy and data utility. That is, it is mathematically impossible to maintain the full data value without privacy leakage risks. The effectiveness of creating privacy-preserving synthetic data has been demonstrated in some cases. For example, Barak et al.~\cite{Barak2007PAC} constructed the desired data by adding noise to the Fourier coefficients so that the original data could not be easily re-identified. Cunningham et al.~\cite{Cunningham2021PPSLD} generated synthetic location data from real locations to protect each individual's existence and true location in the original dataset. By introducing privacy-preserving synthetic data in recommendation, the centralized recommendation models can explicitly utilize synthetic data because the sensitive information has been removed from the original data. Besides, the synthetic data can also be released to the public or shared with organizations depending upon users' privacy requirements and consent. 

Motivated by the above discussion, in this paper, we present a User Privacy Controllable Synthetic Data Generation model (UPC-SDG for short), which generates privacy-preserving synthetic data from the original users' interaction data that incorporates users' privacy preferences. The proposed UPC-SDG model can provide a privacy guarantee for the original data at both data level and item level. To be more specific, at the data level, we design a selection module to select the items used to generate the synthetic item. To maximize the utility of synthetic data, we employ an attention mechanism to estimate the contribution of each purchased item to the user's preference. The items that contribute less to user preference are considered to be the selected items. At the item level, a synthetic item generation module is proposed to generate the synthetic item using the selected item while considering the user's preferences. To maximize the utility of synthetic items while providing necessary privacy protection for the original items, we present a privacy-utility trade-off strategy to optimize the item generation process. Extensive experiments on three real-world datasets have been carried out to demonstrate the effectiveness of our model. We have released the code and relevant parameter settings to facilitate repeatability as well as further research~\footnote{https://github.com/HuilinChenJN/UPC\_SDG.}.

In summary, the main contributions of this work are summarized as follows:

\begin{itemize}
\item We highlight the limitations of existing recommendation systems from the privacy perspective. Motivated by this, we propose a novel UPC-SDG model, which generates privacy-preserving synthetic data from the original data under users' privacy preferences.

\item At the data level, we propose a selection module to select the items which contribute less to the user's preference. At the item level, a synthetic item generation module is developed to create the corresponding synthetic item.

\item To maximize the utility of the synthetic item while providing necessary privacy protection for the original item, we present a privacy-utility trade-off strategy to optimize the synthetic item generation process.

\item We have conducted extensive experiments on three real-world datasets to verify our model and perform comprehensive ablation studies to verify the key assumptions of our model. Experimental results demonstrate the effectiveness of our approach. 
\end{itemize}

\section{Related Work}
\subsection{Collaborative filtering}
Collaborative filtering (CF)~\cite{Rendle2009Bpr,Salak2007PMF,svd++,Hu2008Collaborative,Pan2008One,tkde-fm,Koren2010Factor,hsieh2017collaborative,He2017Neural,cheng2018aspect,wang2019ngcf,liu2019mm,liu2021interest,Cheng2022FLA} has long been recognized as an effective approach for building recommendation systems over the past decades. In model-based CF models, users and items are represented as dense vectors (i.e., embeddings) in the same latent space. An interaction function is used to predict a user's preference for an item based on the learned vectors. Take the Matrix Factorization (MF) approach for example, the user and item embeddings are learned by minimizing the error of re-constructing the user-item interaction matrix, and the dot product is used as the interaction function for prediction. Since this simple idea has achieved great success, many variants of MF have been developed~\cite{Rendle2009Bpr,Salak2007PMF,svd++,Hu2008Collaborative}. The advent of deep learning has accelerated the development of model-based CF techniques. Due to the powerful capability of deep learning, it has been widely applied in recommendation to learn better user and item embeddings~\cite{xue2017deep,Cheng2018A3ncf,cheng2018aspect} or model more complicated interactions between users and items~\cite{He2017Neural}. For example, NeuMF~\cite{He2017Neural} model employs the deep neural network to model the complex interactions between users and items; A$^3$NCF~\cite{Cheng2018A3ncf} learns the user attention from reviews by using the deep learning technique. More recently, Graph Convolution Network (GCN) techniques have also been applied in recommendation and achieved great success~\cite{berg2019gcmc,wang2019ngcf,wei2019hashtag,Wei2019GRCN,Liu2020A2GCN,liu2021interest}. The advantage of GCN-based recommendation models is attributed to their capability of explicitly modeling high-order proximity between users and items. For example, NGCF~\cite{wang2019ngcf} exploits high-order proximity by propagating embeddings on the user-item interaction graph; IMP-GCN~\cite{liu2021interest} learns user and item embeddings by performing high-order graph convolution inside sub-graphs constructed by leveraging users' interests.

All of the methods mentioned above learn the representations of users and items heavily relying on the historical interaction data collected from users in various recommendation scenarios. And their performance suffers when the interactions are insufficient in quantity or when users refuse to share their data due to the sensitive information leakage concern. The use of privacy-preserving synthetic data provides an alternative to combat this problem in recommendation systems. 
\subsection{Privacy-Preserving Synthetic Data}
Open release and free exchange of data would be beneficial for research and development but is not always feasible for sensitive data with privacy implications such as clinical and genomics data.  

Solutions have been proposed in two broad categories to address this privacy challenge. In the first category, the data anonymization-based approaches~\cite{Sweeney2002MPP,Barak2007PAC} try to use various definitions to sanitize data so that it cannot be easily re-identified. For example, Barak et al.~\cite{Barak2007PAC} constructed the desired data from the original data via adding noise to the Fourier coefficients. Although these approaches have some important use cases, they are not usually based on rigorous privacy definitions that can withstand various types of re-identification attacks. In the second category, synthetic data generation approaches have been proposed to generate realistic synthetic data using rigorous differential privacy definitions~\cite{Acs2017DPMGNN,Bindschaedler2017PDPPDS,Cunningham2021PPSLD}. To maximize the utility of the data, the distribution of the generated synthetic data should be as close as possible to that of the original dataset. Still, it should not contain synthetic examples that are too close to real data instances, as the privacy of the original dataset could be compromised. In particular, Acs et al.~\cite{Acs2017DPMGNN} first clustered the original datasets into k clusters with private kernel k-means. Afterward, the generative neural network is adopted to produce synthetic data for each cluster. Bindschaedler et al.~\cite{Bindschaedler2017PDPPDS} introduced the idea of plausible deniability instead of adding noise to the generative model directly. A privacy threshold ensures plausible deniability in releasing synthetic data. Here, an adversary cannot tell whether a particular input belongs to the original data by observing synthetic records. Cunningham et al.~\cite{Cunningham2021PPSLD} presented two methods with high practical utility for generating synthetic location data from real locations, both of which protect each individual's existence and true location in the original dataset.

Although these approaches have been shown to work in some cases, there is still a challenge in generating privacy-preserving synthetic data in recommendation scenarios. In this work, we present a privacy-preserving synthetic data generation model for recommendation.

\subsection{Generative Models in Recommendation}

Generative models have led to rapid development of the deep learning areas, including image~\cite{Zhangnan2019tryon} and video~\cite{Hehefan2021Motion} generation. Numerous studies have also demonstrated the potential of generative techniques to improve recommendation systems by tackling these systems' challenges.

Despite existing recommendation models' pervasiveness and strong performance, recommendation systems still suffer from two main problems: data noise and data sparsity. As an extrinsic problem~\cite{RS2021problem}, data noise stems from the casual, malicious, and uninformative feedback in the training data collected from users in different recommendation scenarios. For example, users might purchase products out of their interests. And, this data noise is also collected and leveraged into the training procedure of the recommendation model. Besides, the randomly selected negative samples would also mislead the recommendation models during the training. All in all, the recommendation results could be manipulated by injecting the noisy data into the RS. Several solutions have been proposed for the data noise problem~\cite{He2018Adversarial,Wang2018NMRN-GAN,Cai2018GAN-HBNR,Tong2019CollaborativeGA,Yuan2019ACAE,Fan2019DASO}. 
For instance, He et al.~\cite{He2018Adversarial} verified the effectiveness of adding adversarial perturbations into a recommendation system, and they proposed an adversarial personalized ranking model to enhance model generalizability. 
Compared to the data noise problem, data sparsity is an intrinsic problem~\cite{RS2021problem}. It occurs because each user generally only consumes a small proportion of the available items. Existing CF models typically rely on the historical interactive information between users and items when capturing users' interests. When a vast quantity of the data is missing, the CF models commonly fail to capture users' preferences and often provide inaccurate recommendations. To alleviate the data sparsity problem, numerous generative models have been developed to improve the performance of recommendation systems by augmenting them with missing interactive information~\cite{Chae2018CFGANAG,Wei2018PLASTIC,Wang2019EnhancingCF,Tran2019MASR,Rafailidis2019ATR}. For example, Chae et al.~\cite{Chae2018CFGANAG} generated the purchase vectors of users rather than item IDs; Wang et al.~\cite{Wang2019EnhancingCF} generated interactions for different recommendation tasks under different auxiliary information conditions. 

In this work, we focus on the privacy issue in recommendation systems. We present a user privacy controllable synthetic data generation model, which aims to remove sensitive information based on the users' privacy preferences while maximizing the utility of the original data. 

\section{The Proposed Model}
\subsection{Preliminaries}
\subsubsection{\textbf{Problem Setting and Notation.}}
Before describing our model, we would like to introduce the problem setting first. 
Given a dataset containing the real user-item interaction data, we aim to generate the synthetic interaction data under users' privacy preferences~\footnote{Note that all our experiments are conducted in a safe environment.}.
For a user set $\mathcal{U}$ and an item set $\mathcal{I}$, we use $\mathcal{R}^{N_u \times N_i}$ to denote the user-item interaction matrix, in which a nonzero entry $r_{u,i}\in \mathcal{R}$ indicates that user $u \in \mathcal{U}$  has interacted with  item ${i} \in \mathcal{I}$ before; otherwise, the entry is zero. $\mathcal{I}_u$ denotes the set of items that user $u$ has interacted with. Note that the interactions can be either implicit (e.g., click) or explicit (e.g., rating). $N_u$ and $N_i$ are the numbers of users and items, respectively. In our setting, each user and item is assigned a unique ID represented by a trait vector.
Let $\mathbf{p_u} \in \mathds{R}^d$ and $\mathbf{q_i} \in  \mathds{R}^d$ denote the feature vector of the user $u$ and a historical item $i$, respectively. $d$ is the dimensionality of the latent space. 
Therefore, for the historical data $\mathcal{I}_u$ of user $u$, our goal is to generate the synthetic data $\mathcal{V}_u$, in which the items generated by our model are used to replace a certain percentage of the original items. Note that the synthetic users' interaction data contains few or no sensitive information. 

To evaluate the utility of the synthetic data, several recommendation models are employed to model users' preferences over the synthetic dataset. 

\subsubsection{\textbf{Privacy Definitions.}} In this work, we provide both data level and item level privacy protection for the original data by using the following two parameters. 

\textbf{Definition 1 (Replacement Ratio).} The replacement ratio $k$ is defined to evaluate the privacy guarantee for the original data at the data level. It is certain that if more original items are replaced by synthetic items, the synthetic data could provide higher level privacy guarantees for the original data. For the given original data set $\mathcal{I}_u$ and synthetic data set $\mathcal{V}_u$, the replacement ratio $k$ is defined as:
\begin{equation}
k = \frac{|\mathcal{I}_u \land \mathcal{V}_u|}{|\mathcal{I}_u|}.
\end{equation}

\textbf{Definition 2 (Sensitivity).} The sensitivity $\gamma$ is defined to describe the privacy guarantee for the original item at the item level. It is obtained by calculating the relative similarity between the original and synthetic data. Given the original data $i \in \mathcal{I}_u$ and the synthetic data $v \in \mathcal{V}_u$, the relative similarity between the original item $i$ and the synthetic item $v$ is calculated by: 
\begin{equation}
f_{sim} \left(\bm{q}_{i}, \bm{q}_{v}\right) = \frac{\bm{q}_{i}^T \bm{q}_{v} - \min(\bm{q}_{i})}{\bm{q}_{i}^T \bm{q}_{i} - \min(\bm{q}_{i})} ,
\end{equation}
where $\min \left(\cdot \right)$ denotes the cosine similarity between item $i$ and the item which is the most insensitive one in the item set $\mathcal{I}$. 
In our definition, the synthetic item can meet user's privacy requirement in terms of the sensitivity if $f_{sim} (\bm{q}_{i}, \bm{q}_{v}) \leq \gamma $.

Intuitively, different users have different privacy preferences. Therefore, both $ k \in \left( 0, 1 \right) $ and $ \gamma \in \left( 0, 1 \right) $ are utilized as parameters which can be adjusted to change the security level of the synthetic data in our proposed model, and the two parameters are controlled by users based on their personal privacy preferences. 

\subsection{Model Overview}

In this section, we present an overview of the proposed User Privacy Controllable Synthetic Data Generation (short for UPC-SDG) model, which aims to generate privacy-preserving synthetic data based on the users' privacy preferences. 

It should be noted that UPC-SDG is designed for providing the privacy guarantee to the original data while maximizing the utility of the synthetic data from both data level and item level perspectives. 
At the data level, we present a selection module, it needs to select the items which will be replaced by the generated items from the original data. The number of items that need to be replaced by the generated synthetic items depends on the user's privacy requirements, which can be adjusted by setting the replacement ratio $k$. To maximize the utility of the synthetic data at the data level, we present an attention mechanism to estimate all items' contributions to the user's preference. The items that contribute less to the user's preference are selected for replacement by the generated synthetic items.
At the item level,
a generation module is developed to generate the item from all items based on the users' preferences, the users' privacy preferences and the characteristics of the selected items. The generation module aims to generate the items that are appealing to a user while meeting the user's privacy requirements in terms of sensitivity. In other words, the users' privacy preferences are controllable and could be adjusted by setting the sensitivity $\gamma$. In addition, for balancing the privacy and utility between generated items and original items, a privacy-utility trade-off strategy is deployed. The strategy can provide the privacy guarantee of the original item while maximizing the utility of the synthetic item.

\subsection{Synthetic Data Generation} 
The proposed UPC-SGD model aims to provide privacy guarantees while maximizing the utility for the original user interaction data at both the data level and the item level. Next, we detail the model for privacy-preserving synthetic data generation.

\subsubsection{\textbf{Data level.}}
For a user $u$, her privacy is protected if the non-sensitive items replace the sensitive items in the original data. In this work, the number of items that need to be replaced is determined using the replacement ratio $k$. The more items are replaced, the less data leakage risk there is.
To maximize the utility of the synthetic data at the data level, we design a selection module to find the items which contribute less to the user's preference than the other items in the original data.

\textbf{Selection Module.} It has shown that historically interacted items have different contributions to model users' preferences~\cite{He2018Nais,Cheng2022FLA}. Therefore, by employing the representations of items that a user $u$ purchased, the user's preferences can also be presented as follows:

\begin{equation}
\bm{t_u} = \frac{1}{|\mathcal{I}_u|} \sum_{i\in \mathcal{I}_u} a_{ui} \bm{q_i},
\end{equation}
where $a_{ui} \in \bm{A}_{u}$ is a trainable parameter that denotes the attention weight of item $i$ in contributing to user $u$'s preference; $\bm{A}_{u}$ represents the weight set of the purchased items by the user $u$. 
Formally, the attention weight $a_{ui}$ is estimated by an attention mechanism which is described as following:
\begin{equation}
   v_{ui} =\mathbf{h}^T ReLU(\mathbf{W}_1([\mathbf{p}_u :\mathbf{q}_i])+\mathbf{b}_1),
\end{equation}
\begin{equation} \label{eq:tstepsoft}
   a_{ui}=\frac{\exp(v_{ui})}{[\sum_{j'\in\mathcal{I}_u}\exp( v_{ui'})]^\beta},
\end{equation}
where $\bm{W}_1$ and $\bm{b}_1$ are respectively the weight matrix and bias vector that project the input onto a hidden layer, and $\mathbf{h}^T$ is the vector that projects the output of the hidden layer to an output attention weight $v_{ui}$. $\beta$ is the smoothing exponent. It is employed as a hyperparameter which is set in [0, 1]. We use the Rectified Linear Unit (ReLU) as the activation function of the hidden layer.

As both $p_u$ and $t_u$ can capture user's preferences, we employ $p_u$ as the supervision information to learn the attention weight $a_{ui}$. To be specific, We define the following $L_2$ regularizer:
\begin{equation}
\mathcal{L}_{D} =\sum_{u \in \mathcal{U}} \|\bm{f(\bm{t}_u,\theta)}-\bm{p}_u\|^2,
\end{equation}
where we learn a transformation function $f(\cdot)$ that projects $\bm{t}_u$ into the common representation space with $\bm{p}_u$. $\theta$ denotes the parameters of function $f(\cdot)$. Note that we employ a multi-layer perceptron (MLP) with dropout as our transformation function $f(\cdot)$ for its superior representational capacity and ease of training. The parameters of the MLP are trainable. During the training, we minimize $\mathcal{L}_D$ to estimate the contributions of each item to the user preference.
In this way, the attention weights of all purchased items for user $u$ are obtained. With the obtained attention weights, the items that contribute less to the user's preference are selected as the replaced item:
\begin{equation}
\bm{I}_u^r = \{i | i \in \bm{I}_u, a_{ui} \in \bm{A}_u^k \} ,
\end{equation}
where $\bm{A}_u^k$ denotes the $k$ percent of the items, the weights of which are smaller than the others. 

\subsubsection{\textbf{Item level}}
With the selected items $\bm{I}_u^r$, the synthetic items for the user $u$ can be generated. One of the targets of generating the synthetic item is that the generated item can also attract user $u$' attention. In other words, to maximize the utility of the synthetic item, we present a generation module that generates the item considering the users' preferences, the users' privacy preferences and the characteristics of the selected items.

\textbf{Generation Module.}
For generating the synthetic item, we first concatenate the user vector $\bm{p}_u$, the selected item vector $\bm{q}_{i}$ ($i \in \bm{I}_u^r$) and privacy parameter $\gamma$ as an input, then project the input into a latent space:
\begin{equation}
\bm{R_{u,i}} = \bm{W}_{2}[\bm{p}_{u};\bm{q}_{i};\gamma_u] + \bm{b}_{2},
\end{equation}
where $\bm{R_{u,i}}$ is the latent feature of the output. $\bm{W}_2$ and $\bm{b}_2$ are respectively the weight matrix and bias vector. Let $\bm{E}_I \in \mathds{R}^{N_i \times d}$ represents the embeddings of all items, we calculate the similarity between the latent feature $\bm{R_{u,i}}$ and all item embeddings:
\begin{equation}
\bm{h_{u,i}} = \bm{R_{u,i}} {\bm{E}_I}^T.
\end{equation}
The output $\bm{h_{u,i}}$ whose elements represent the similarity of the latent feature $\bm{R_{u,i}}$ with each item embedding, which is going to be learned during training. 
Finally, we estimate the probability distribution over all candidate items. It is formulated as:
\begin{equation}
\bm{y_{u,i}} \sim softmax(\bm{h_{u,i}}), 
\end{equation} 
where $\bm{y_{u,i}}$ represents the probability distribution over all possible items. Intuitively, the item with the highest possibility should be sampled as the generated synthetic item for the selected item. However, there is a crucial problem with sampling that the discrete sampling procedure is non-differentiable. In other words, the training of the generation module is not end-to-end, resulting in an inferior performance. The recently proposed Gumbel-Softmax~\cite{Gumbelsoftmax2017nips,jang2017categorical} can approximate categorical samples by applying a differentiable procedure.
Thus, Gumbel-Max trick is proposed to re-parameterize the sampling operator as below:
\begin{equation}
\bm{y_{u,i}} = one\_hot(argmax_{1 \leq k \leq N_i}(h_{u,i}^k + g_{u,i}^k)),
\end{equation} 
where $h_{u,i}^k \in \bm{h_{u,i}}$ and $g_{u,i}^k \in \bm{g_{u,i}}$. The $g_{u,i}^k$ is an independent variable and follow a Gumbel distribution $g_{u,i}^k = -\log_{} {(-\log_{} {(\mu)})}$ whith $\mu \sim Uniform(0,1)$. However, the ``one\_hot" and ``argmax” operations are still non-differentiable. To deal with this problem, we further approximate them by softmax, which yields
\begin{equation}
\label{eq:gumbel-softmax}
\bm{\tilde{y_{u,i}}} = softmax(\frac{1}{\tau}(\bm{h_{u,i}} + \bm{g_{u,i}})).
\end{equation} 
Note that $\tau > 0$ is a tunable parameter called temperature to control the softness. As $\bm{\tilde{y_{u,i}}}$ is differentiable with respect to $\bm{h_{u,i}}$, it can be used instead of $\bm{y_{u,i}}$.

After the scores $\bm{y_{u,i}}$ for candidate items are obtained, we sample the item not consumed by the user $u$ with the Gumbel-Softmax as shown in Eq.~\ref{eq:gumbel-softmax}. Finally, the generated item is delivered to the loss function to help optimize the generation process. If the generated synthetic item has properties similar to that of the selected item, there is still the risk of privacy leakage. Hence, for providing privacy guarantees while maximizing the utility of the synthetic item, we propose a privacy-utility trade-off strategy at the item level.  

\textbf{Privacy-utility Trade-off Strategy.}
To provide privacy guarantees at the item level, we employ a privacy regularizer to constrain the differences of relative similarity between the selected original item and the generated synthetic item. The privacy regularizer is defined as:
\begin{equation}
\mathcal{L}_{s} = \sum_{(u,i,v)}[f_{sim}(\bm{q}_{i}, \bm{q}_v) - \gamma_u]_+.
\end{equation} 
where  $[ z ]_{+} = max( z,0)$ denotes the standard hinge loss. For a user $u$, the sensitivity $\gamma_u$ is adopted as a safety margin that the similarity between the selected original item and the generated synthetic item is tolerated within a certain range.

Inspired by the recommendation approach in~\cite{He2017Neural}, we assume that a user $u$ would give the synthetic item a higher score if she prefers this item. To maximize the utility of synthetic items, we aim to generate the items (i.e., $v$) that can attract the user's attention. To achieve the goal, we construct a user-item pair of $\{u, v\}$ with a generated interaction between user $u$ and item $v$, and define the following utility regularizer:

\begin{equation}
\mathcal{L}_{g} = \sum_{(\mathbf{u}, \mathbf{v})} -\ln\sigma(\bm{p}_u^T \bm{q}_v).
\end{equation} 
The final loss function of the privacy-utility trade-off strategy is formulated as:
\begin{equation}
\mathcal{L}_{I} =\lambda_s\mathcal{L}_s + \lambda_g\mathcal{L}_g,
\end{equation} 
where $\lambda_s$ and $\lambda_g$ are hyperparameters that control the weights of the privacy regularizer and utility regularizer, respectively. The loss function of the generation model can be interpreted as the model trying to produce such items that the user prefers.
Finally, the proposed privacy-utility trade-off strategy finely balances the privacy and utility of the synthetic item at the item level.

\subsection{Optimization}

In this work, we focus on generating privacy-preserving synthetic data for the Top-$N$ recommendation. With the above consideration of the selection and generation module, the final objective function for optimizing the parameters of UPC-SGD is:
\begin{equation}
\mathcal{L} = \mathcal{L}_D + \mathcal{L}_I. 
\end{equation}
The mini-batch Adam~\cite{kingma2014adam} optimizer is used for optimizing the generation model. And the model parameters are updated by using the gradients of the loss function.
\section{Experiment}
To evaluate the effectiveness of our proposed UPC-SDG model, we perform extensive experiments on three publicly real-world datasets. Through the experiments, we attempt to answer the following research questions.
\begin{itemize}
  \item{\textbf{RQ1:}} How is the existing recommendation models' performance on the synthetic datasets generated under different privacy settings?
  \item{\textbf{RQ2:}} Does the synthetic data remove the sensitive information from the original data?
  \item{\textbf{RQ3:}} How do different components of our model affect the performance?
\end{itemize}
 
Next, we introduce the experimental setup and then sequentially report and analyze the experimental results to answer the above questions.
 
\subsection{Experimental Setup}
\subsubsection{\textbf{Datasets}}
To evaluate the effectiveness of UPC-SDG, we conducted experiments for the generation task on three public real-world datasets: \emph{Office}, \emph{Clothing}, and \emph{Gowalla}.  

The first two datasets are from the public Amazon review dataset~\footnote{http://jmcauley.ucsd.edu/data/amazon.}, widely used for recommendation evaluation in previous studies~\cite{liu2019mm} and adopted for experiments in this work.
The third dataset is a check-in dataset collected from Gowalla~\cite{wang2019ngcf}, where users share their locations by checking in.
Following the settings in previous studies~\cite{liu2019mm,wang2019ngcf}, we pre-process all the datasets to keep only the user and items with at least 10 interactions. The basic statistics of the three datasets are shown in Table~\ref{tab:data}. We can see that the three datasets are of different sizes and sparsity. For example, the \emph{Office} dataset is denser, and the \emph{Gowalla} dataset is larger than other datasets. The diversity of datasets helps analyze the performance of our method in different situations.

In this work, we focus on the privacy-preserving synthetic data generation task, which aims to generate interaction data for each user under the user's privacy preference. Besides, we verify the utility of the generated data based on the recommendation task.
Thus, for each dataset, we randomly split each dataset into training, validation, and testing sets with a ratio of 80:10:10 for each user. 
The observed user-item interactions in the original three datasets were treated as the input feature in the generation task.

For the recommendation task, the experiments are conducted over the generated datasets. The observed user-item interactions in the generated datasets were treated as positive instances. In addition, for the recommendation methods which adopt the pairwise learning strategy, we randomly sample a negative instance that the user did not consume before, to pair with each positive instance.

\begin{table}
	\caption{ Basic statistics of the experimental datasets.}
	\vspace{-0.0cm}
	\label{tab:data}
	\begin{tabular}{ccccl}
		\toprule
		Dataset&\#user&\#item&\#interactions&sparsity\\
		\midrule
		Office & 4,874& 2,405 & 52,957 & 99.55\%\\
		Clothing & 18,209& 17,317 & 150,889 &99.95\%\\
		Gowalla & 29,858 & 40,981 & 1,027,370 &99.91\%\\ 
		\bottomrule
	\end{tabular}
	\vspace{0.0cm}
\end{table}
\begin{table*}[ht]
	\caption{Performance comparison of recommendation models over the original and generated datasets. Specifically, we use the proposed UPC-SDG model to separately generate two synthetic datasets for each original dataset under two privacy settings. Such as $Office^{-}$ is generated with $k=0.8$ and $\gamma=0.1$; and $Office^{+}$ is generated with $k=0.2$ and $\gamma=0.9$. In addition, ``-" indicates that the recommendation model has the same performance over the generated datasets as the original datasets.}
\begin{tabular}{l|ccc|ccc|ccc|ccc}
\hline
\multicolumn{1}{l|}{Models}  & \multicolumn{3}{c|}{Random}                     & \multicolumn{3}{c|}{BRPMF}                       & \multicolumn{3}{c|}{NeuMF}                       & \multicolumn{3}{c}{LightGCN}                  \\ \hline
\multicolumn{1}{l|}{Metrics} & PR & Recall  & \multicolumn{1}{c|}{NDCG} & PR & Recall & \multicolumn{1}{c|}{NDCG} & PR & Recall & \multicolumn{1}{c|}{NDCG}   & PR                   & Recall & NDCG   \\ \hline\hline

Office        & 0.0021  & 0.0118  & 0.0054  & 0.0215  & 0.1254  & 0.0740  & 0.0126  & 0.0605  & 0.0710  & 0.0223  & 0.1323  & 0.0752 \\
Clothing      & 0.0001  & 0.0019  & 0.0006  & 0.0039  & 0.0526  & 0.0243  & 0.0120  & 0.0668  & 0.0303  & 0.0043  & 0.0692  & 0.0312 \\
Gowalla       & 0.0005  & 0.0012  & 0.0008  & 0.0475  & 0.1565  & 0.1277  & 0.0285  & 0.1017  & 0.1548  & 0.0523  & 0.1823  & 0.1555 \\
\hline\hline
Office$^-$    & -       & -       & -       & 0.0032  & 0.0176  & 0.0080  & 0.0058  & 0.0316  & 0.0374  & 0.0046  & 0.0235  & 0.0125 \\
Clothing$^-$  & -       & -       & -       & 0.0001  & 0.0023  & 0.0026  & 0.0007  & 0.0109  & 0.0044  & 0.0020  & 0.0345  & 0.0148 \\
Gowalla$^-$   & -       & -       & -       & 0.0035  & 0.0121  & 0.0073  & 0.0030  & 0.0088  & 0.0213  & 0.0047  & 0.0148  & 0.0100 \\
\hline\hline
Office$^+$    & -       & -       & -       & 0.0166  & 0.0878  & 0.0506  & 0.0121  & 0.0580  & 0.0702  & 0.0148  & 0.0839  & 0.0496 \\
Clothing$^+$  & -       & -       & -       & 0.0025  & 0.0422  & 0.0187  & 0.0110  & 0.0553  & 0.0245  & 0.0038  & 0.0639  & 0.0281 \\
Gowalla$^+$   & -       & -       & -       & 0.0305  & 0.1051  & 0.0740  & 0.0185  & 0.0736  & 0.0870  & 0.0353  & 0.1257  & 0.0812 \\
\hline
\end{tabular}
\vspace{0.0cm}
\label{tab:performance1}
\end{table*}
\subsubsection{\textbf{Evaluation Metrics}}
We evaluate the utility of our generated interaction data on the top-$n$ recommendation task.
For each user in the test set, we treat all the items that the user did not interact with as negative items. Three widely used metrics for the recommendation are used in our evaluation: \emph{Precision}~(PR), \emph{Recall} and \emph{Normalized Discounted Cumulative Gain}~(NDCG)~\cite{he2015trirank}.
For each metric, the performance is computed based on the top 20 results. Notice that the reported results are the average values across all the testing users. 

\subsubsection{\textbf{Recommendation Models}}
This work aims to generate privacy-preserving synthetic data for users in recommendation systems. Therefore, we explore the utility of the generated interaction data with the following recommendation models, such as Random~\cite{Cheng2022FLA}, BPRMF~\cite{Rendle2009Bpr}, NeuMF~\cite{He2017Neural} and LightGCN~\cite{He2020lightgcn}.

\begin{itemize}
        \item \textbf{Random} recommends items randomly to users. 
        
        \item \textbf{BPRMF}~\cite{Rendle2009Bpr} is a classic MF-based method for top-$n$ recommendation. It only utilizes the user-item interaction data and uses the Bayesian personalized ranking loss to learn user and item embeddings.
        
        \item \textbf{NeuMF}~\cite{He2017Neural} is a neural collaborative filtering method, which employs multiple hidden layers above the element-wise to capture the non-linear interactions between users and items. 
        
        \item \textbf{LightGCN}~\cite{He2020lightgcn} is a simplified GCN-based recommendation model, which explicitly encodes the collaborative signal of high-order neighbors by performing embedding propagation in the user-item bipartite graph.
\end{itemize}

In the experiments on recommendation tasks, for all the recommendation models, we use the code that their authors released for evaluation. And, we put a lot of effort to tune the hyperparameters of these models and reported their best performance.

\subsubsection{\textbf{Experimental Settings}}
Our proposed UPC-SDG model is implemented with PyTorch~\footnote{https://pytorch.org.} and its key parameters are carefully tuned. 
The pre-trained user and item embeddings~\footnote{The pre-trained user and item embeddings are learned with the BPRMF model over the training set of each dataset.} are adopted as user and item features in our model, and the embedding size is fixed to 64. We optimized our method with Adam~\cite{kingma2014adam} and the learning rate is searched in the range of $\{ 1e^{-1}, 1e^{-2}, \cdots, 1e^{-5}\}$. 
Besides, to verify the controllability of the selection module and the generation module of UPC-SDG, we generate various synthetic data for each user with different privacy settings by adjusting the replacement ratio $k \in \{0.2, 0,4, 0,6, 0.8\}$ and the sensitivity $\gamma \in \{0.1, 0,3, 0,5, 0,7, 0.9\}$. 
For balancing the effect of the privacy regularizer and utility regularizer, we set $\lambda_s = 3$ and $\lambda_g = 1$ for \emph{Office} and \emph{Gowalla}, and set $\lambda_s = 5$ and $\lambda_g = 7$ for \emph{Clothing} in generation task.

For the recommendation task, the embeddings size is set to 64 for all recommendation models. We used the default mini-batch size of 2048 for all datasets. The $L_2$ regularization coefficient is searched in the range of $\{ 1e^{-1}, 1e^{-2}, \cdots, 1e^{-6}\}$. All the recommendation models have also been carefully tuned for a fair comparison. In addition, the early stopping and validation strategies are kept the same as those in LightGCN~\cite{He2020lightgcn}.

\subsection{Utility Analysis (RQ1)}
\subsubsection{\textbf{Performance Comparison}}

The results of existing recommendation models over the original datasets (\emph{Office},\emph{Clothing},\emph{Gowalla}) and the datasets generated based on the three datasets under different privacy settings are reported in Table~\ref{tab:performance1} in terms of PR@20, Recall@20 and NDCG@20. Specifically, we use the proposed UPC-SDG model to generate two synthetic datasets separately based on each dataset under two privacy settings. For example, $Office^{-}$ is generated with $k=0.8$, $\gamma=0.1$; and $Office^{+}$ is generated with $k=0.2$, $\gamma=0.9$. Similarly, the synthetic datasets $Clothing^{-}$, $Clothing^{+}$, $Gowalla^{-}$ and $Gowalla^{-}$ are obtained.

We firstly focused on the performance of the recommendation models over the original datasets. Specifically, the performance of Random is poor as it randomly recommends the items to users. Benefiting from the technique of matrix factorization, BPRMF can achieve better performance than the random method. The reason is that the user and item representations are learned by exploiting the interaction data. NeuMF obtains better performance over BPRMF, because it utilizes neural networks to model user-item interactions. LightGCN outperforms both BPRMF and NeuMF by a large margin as it explicitly leverages the high-order connectivities between users and items.

The second and third blocks are the performance of the recommendation models over the generated datasets~\footnote{Random has similar performance over both the original and generated datasets.}. $Office^{-}$, $Clothing^{-}$ and $Gowalla^{-}$ are generated under the higher privacy level ($k=0.8$, $\gamma=0.1$). As we can see, the recommendation models can still model users' preferences over these datasets. On the contrary, $Office^{+}$, $Clothing^{+}$ and $Gowalla^{+}$ are generated under the lower privacy level ($k=0.2$, $\gamma=0.9$). The recommendation models can achieve comparable performance with them over the original datasets. The experimental results demonstrate the utility of the generated datasets.

\subsubsection{\textbf{Effect of Privacy Preference}}
\begin{figure*}[ht]
	\centering
	\subfloat[PR on Office]{\includegraphics[width=0.33\linewidth]{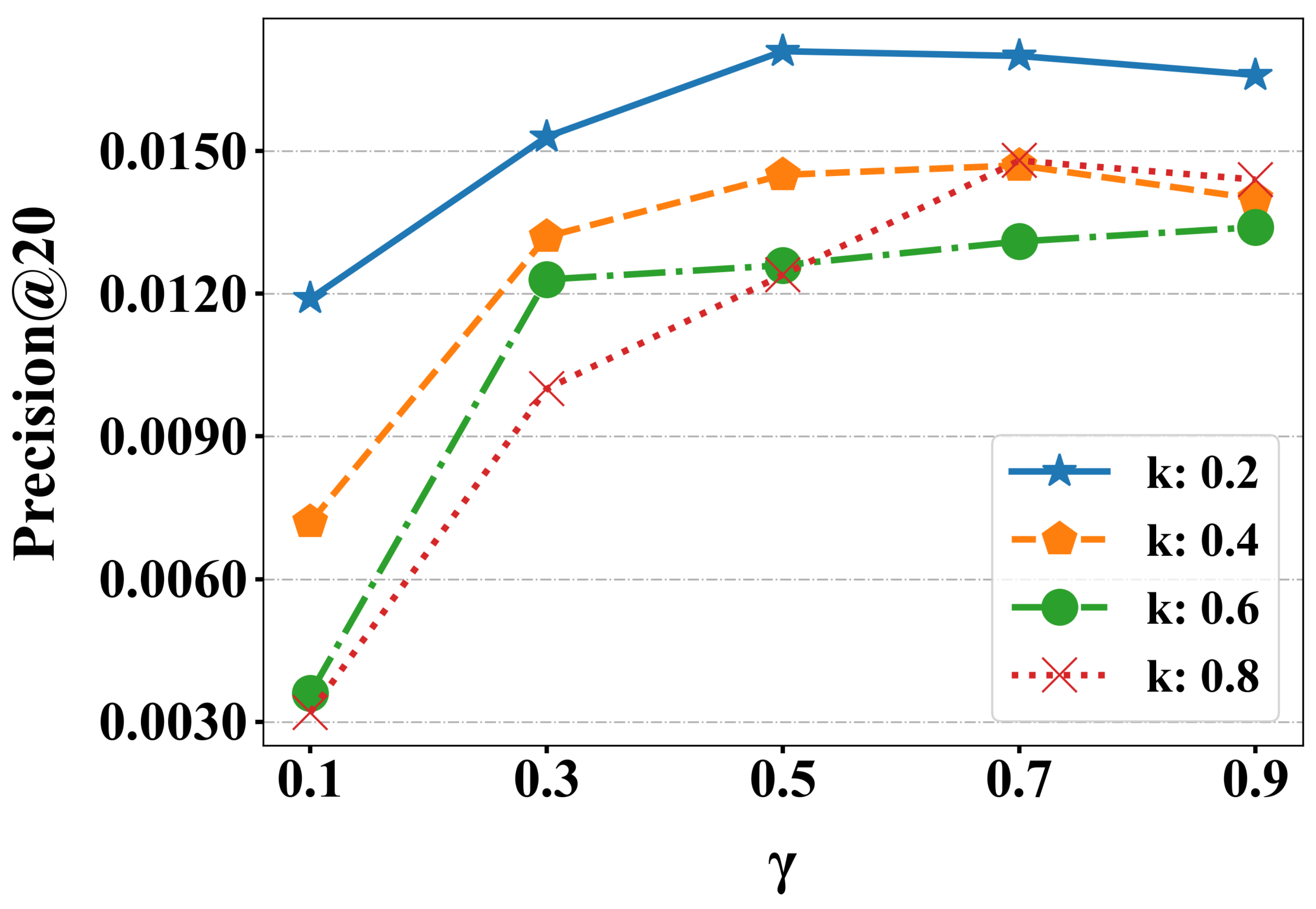}}
	\subfloat[Recall on Office]{\includegraphics[width=0.33\linewidth]{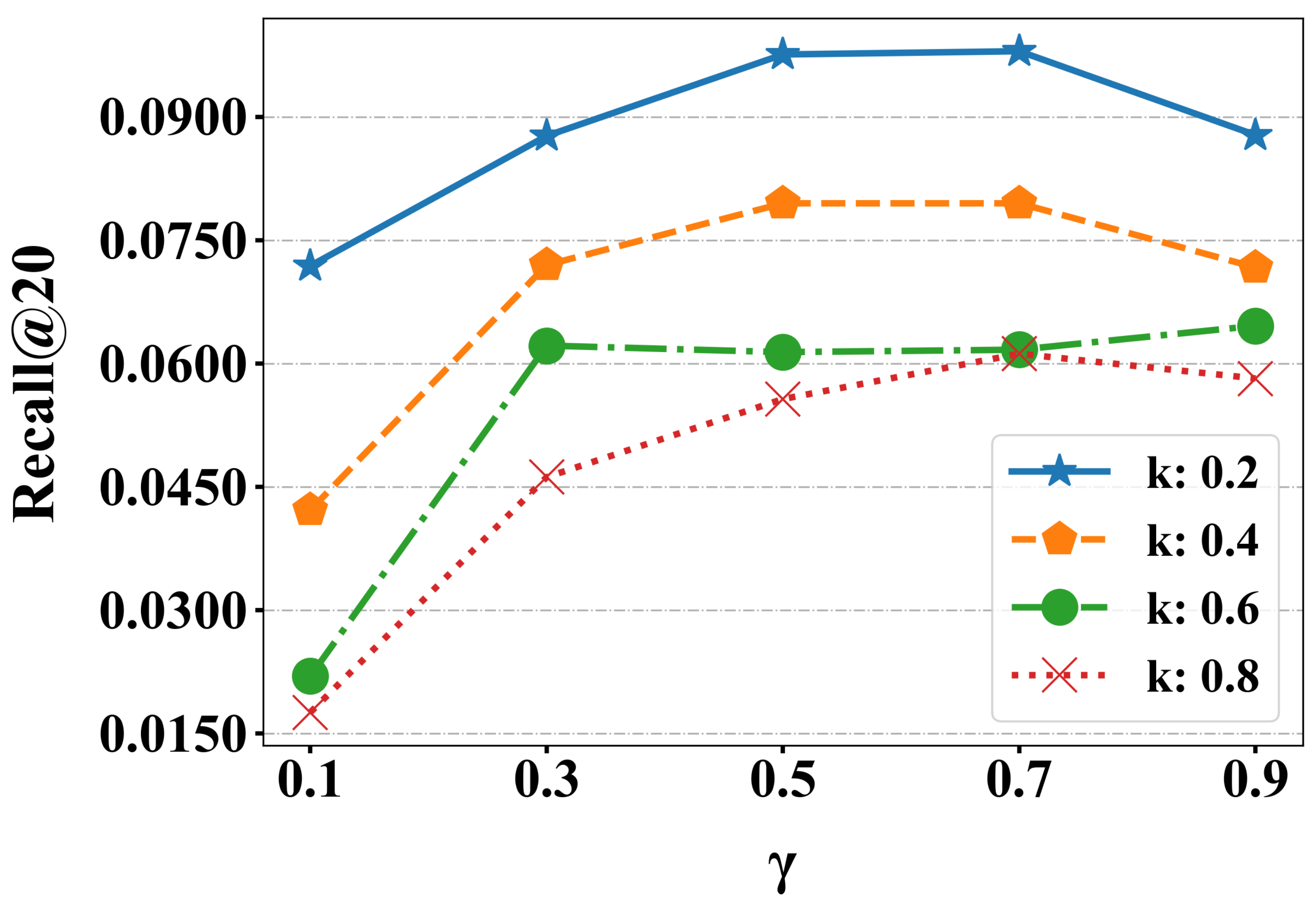}}
	\subfloat[NDCG on Office]{\includegraphics[width=0.33\linewidth]{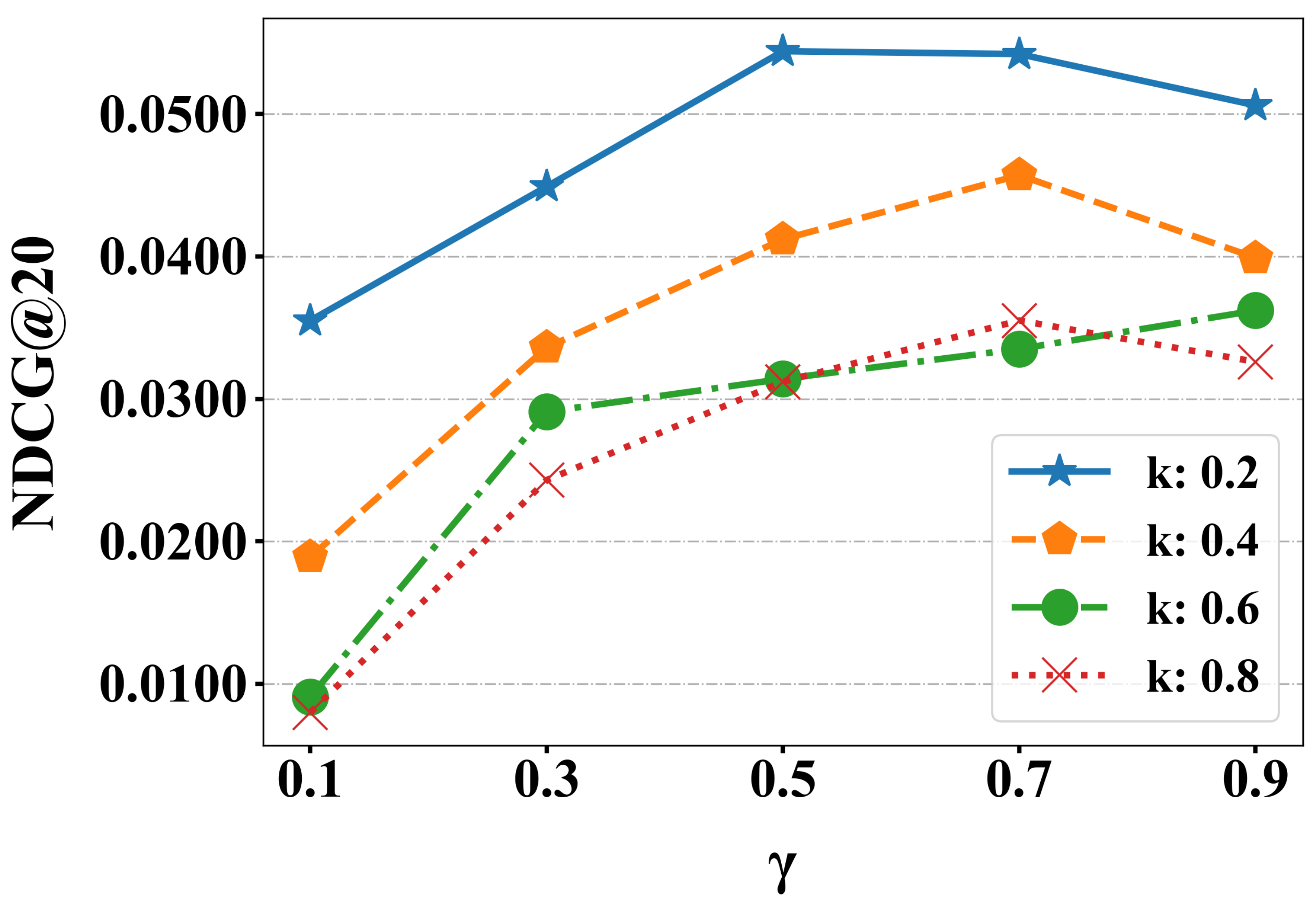}}
	\newline
	\subfloat[PR on Clothing]{\includegraphics[width=0.33\linewidth]{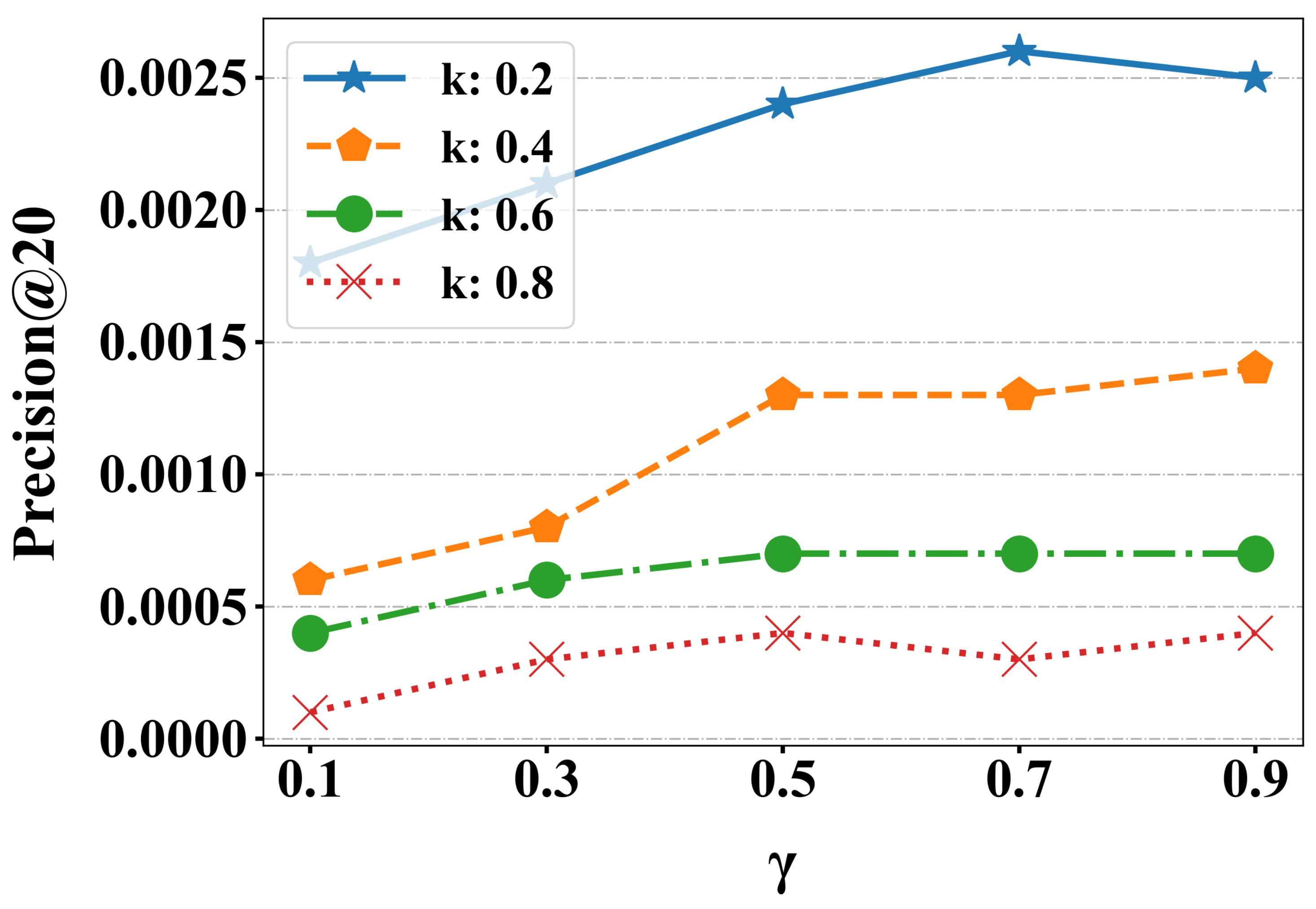}}
	\subfloat[Recall on Clothing]{\includegraphics[width=0.33\linewidth]{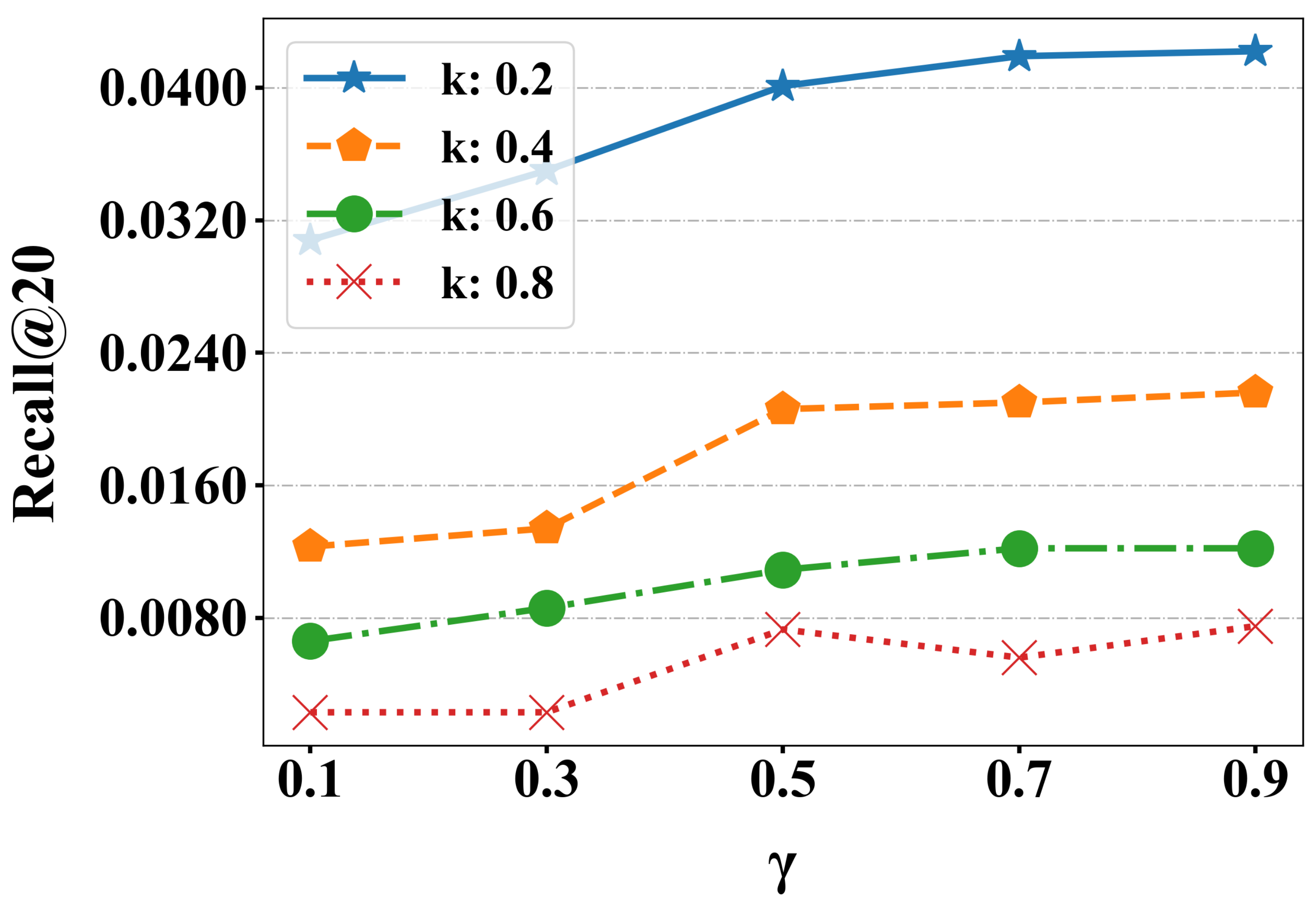}}
	\subfloat[NDCG on Clothing]{\includegraphics[width=0.33\linewidth]{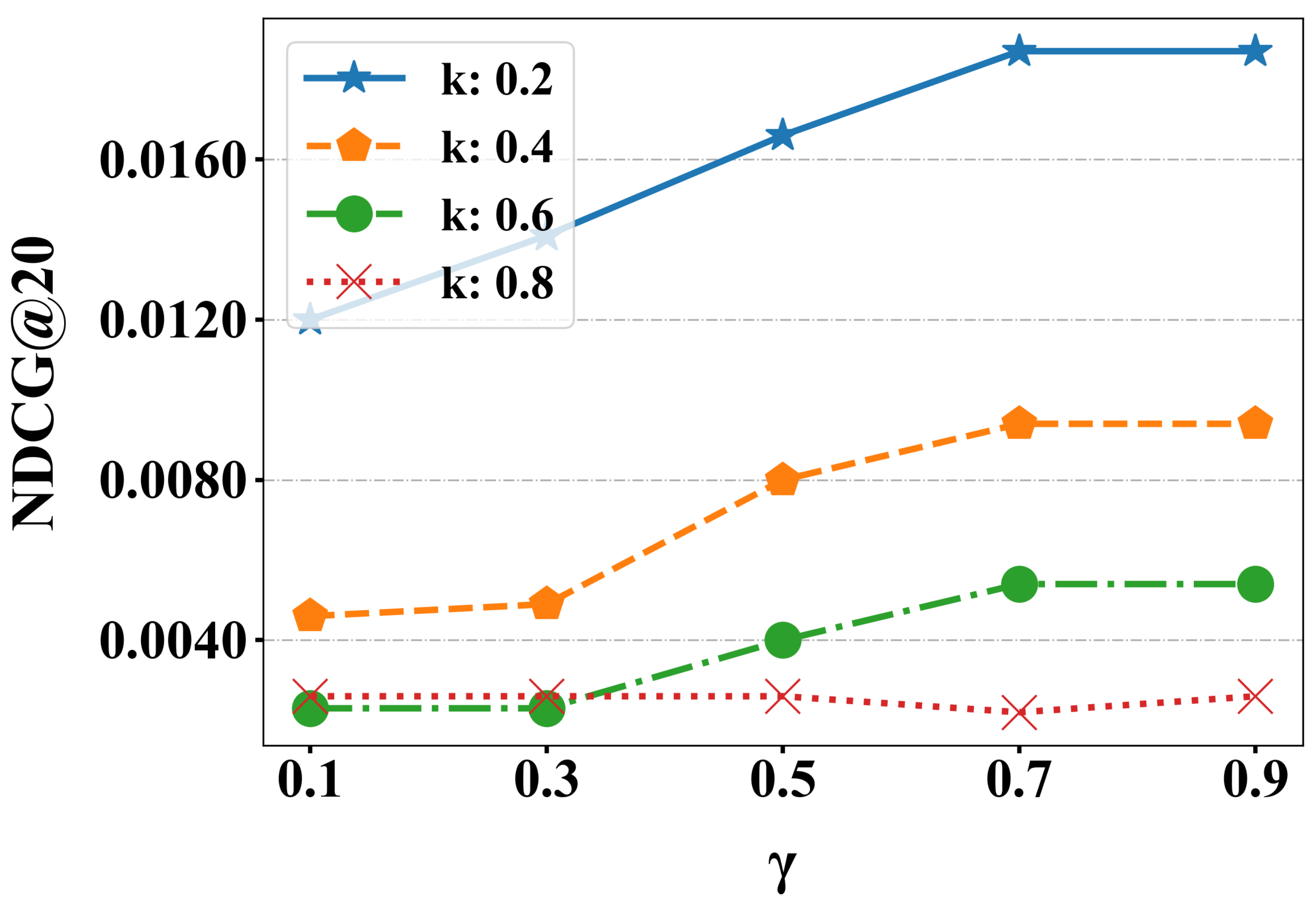}}
	\newline
	\subfloat[PR on Gowalla]{\includegraphics[width=0.33\linewidth]{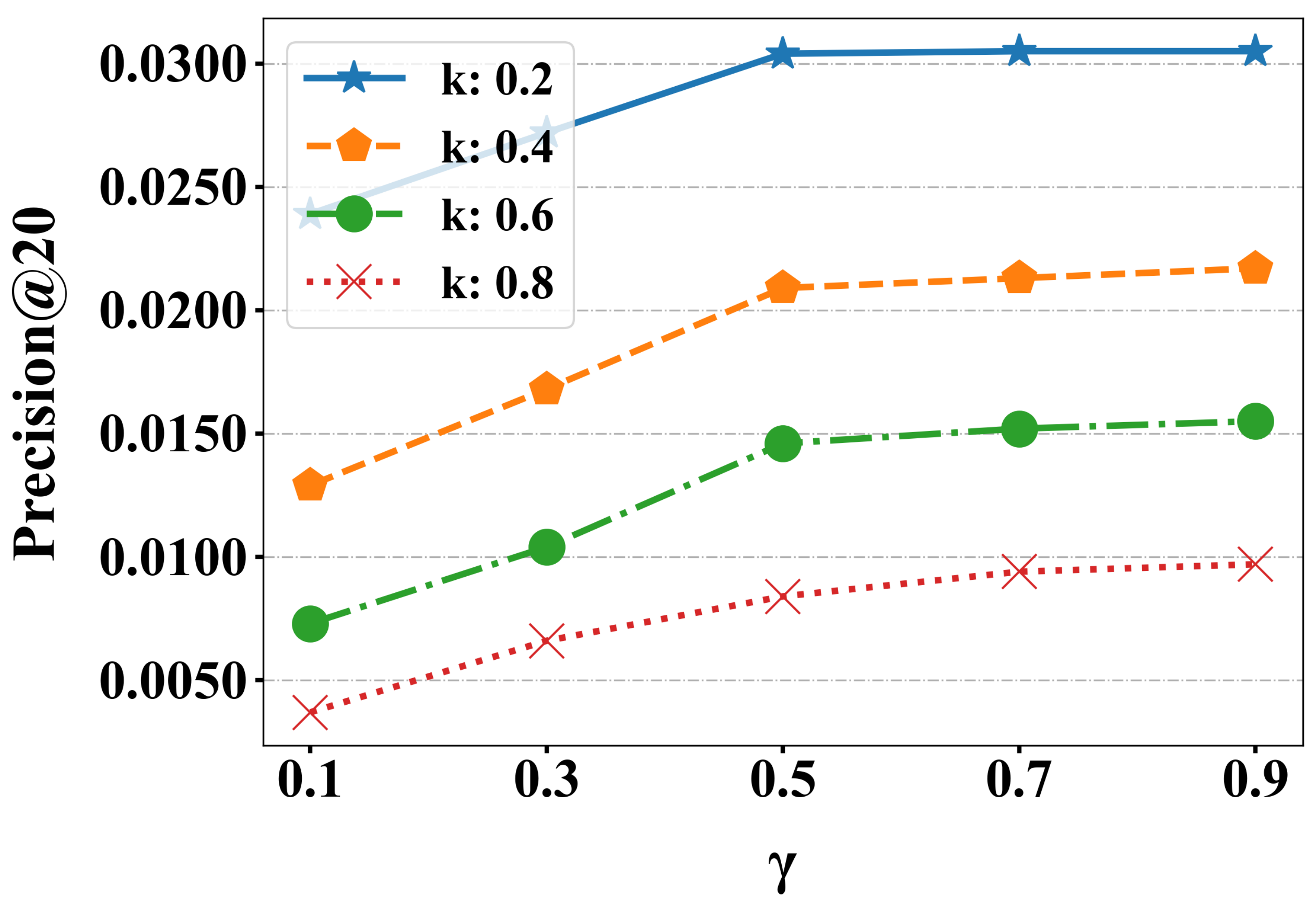}}
	\subfloat[Recall on Gowalla]{\includegraphics[width=0.33\linewidth]{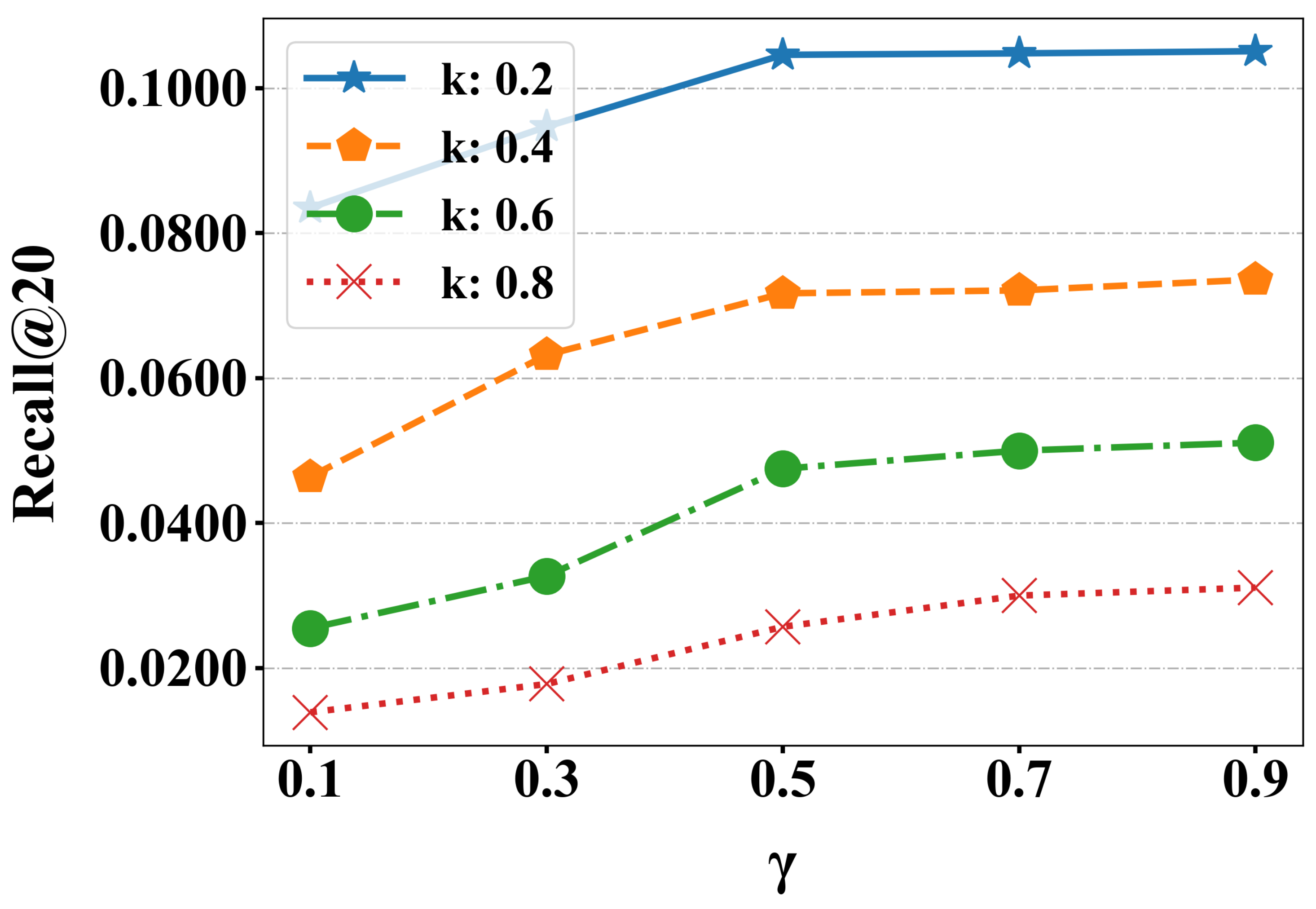}}
	\subfloat[NDCG on Gowalla]{\includegraphics[width=0.33\linewidth]{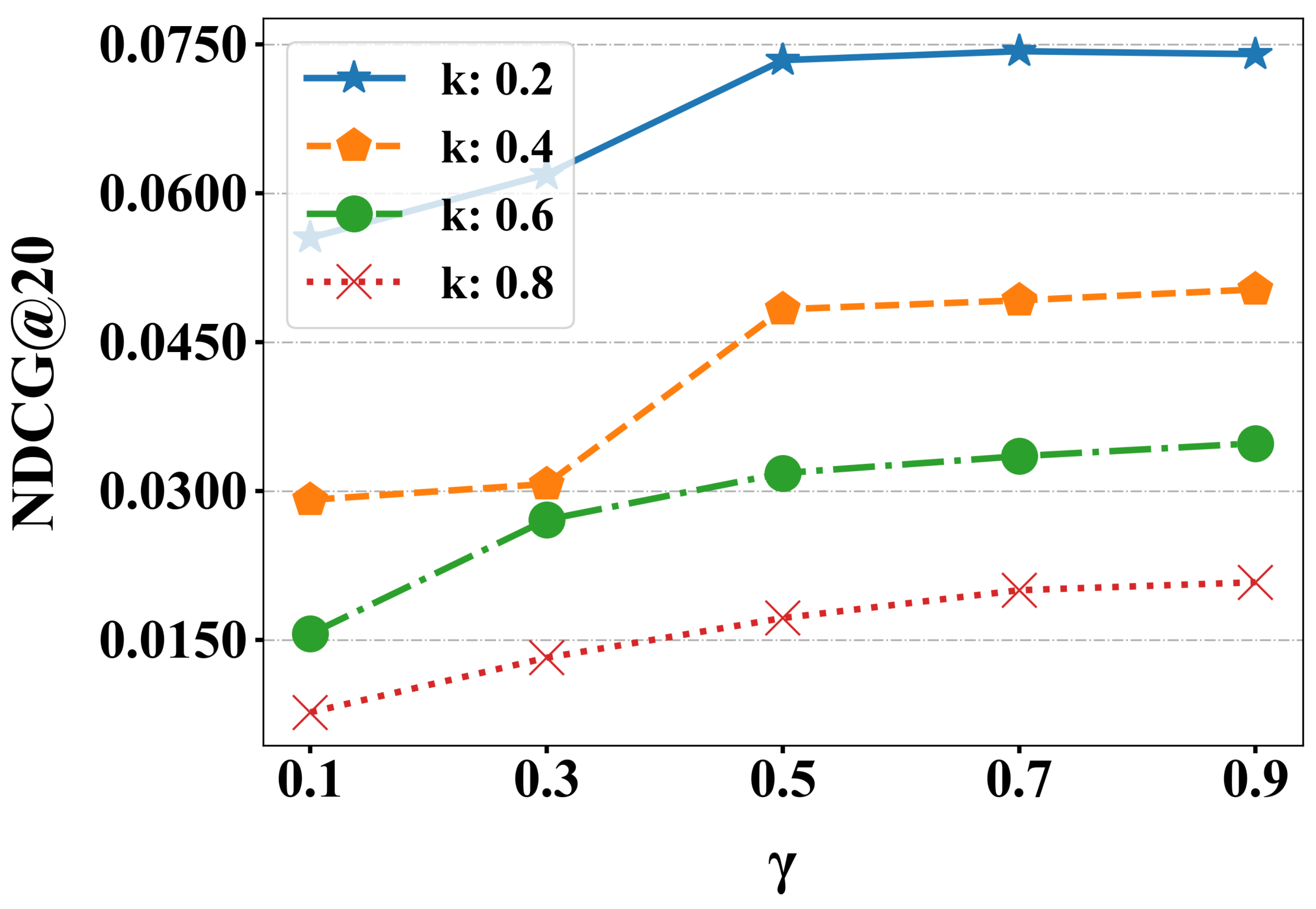}}
	\newline
	\vspace{0.0cm}
	\caption{Performance comparison of BPRMF over the generated datasets of different privacy settings. $k$ and $\gamma$ represent replacement ratio and sensitivity, respectively.}
	\label{fig:similarity}
\end{figure*}

To investigate the utility of the generated datasets under different privacy preferences, we generated various datasets based on the \emph{Office}, \emph{Clothing} and \emph{Gowalla} under different privacy settings. Fig.~\ref{fig:similarity} shows the performance of the BPRMF over various generated datasets. From the results, we obtained some interesting observations:

\begin{itemize}
\item The results indicate that with the increase of replacement ratio $k$ and the decrease of sensitivity $\gamma$, the performance of the recommendation model decreases as expected. Moreover, the generated dataset can still  be used to model users' preferences, even when the replacement ratio $k$ is set as 0.8 and the sensitivity $\gamma$ is set as 0.1. This also demonstrates the effectiveness of our model in retaining the utility of synthetic data. Note that the performance of BPRMF drops when $\gamma$ is set as 0.9. The reason might be that the generated item is very similar to the items in the original user interaction data. Thus, the diversity of the items in each user's interaction data is reduced, which leads to sub-optimal performance.
\item The best trade-off point is different for users in different recommendation scenarios. From the perspective of Sensitive $\gamma$, the best trade-off points are 0.5, 0.7, 0.5 on \emph{Office}, \emph{Clothing}, and \emph{Gowalla}, respectively~\footnote{This depends on whether the recommendation performance will degrade rapidly compared with the recommendation performance over the generated datasets of other settings.}. In other words, users need to give up more privacy in exchange for recommendation performance on \emph{Clothing}. 
\item The sensitivity $\gamma$ has different impact on different datasets. From the results, the impact of sensitivity adjustment on \emph{Gowalla} is lower than in the other datasets. That might be because the designed generation model introduces numerous parameters, and the model is difficult to train on smaller datasets.
\item The replacement ratio $k$ has different impact on different datasets. From the results, the impact of replacement adjustment on  \emph{Office} is lower than in the other datasets. The reason is that $Office$ is denser than other datasets. With rich interactions, users' preferences could still be modeled well even if some of the interaction data are replaced by synthetic data.
\end{itemize}

Overall, the proposed UPC-SDG model can generate  privacy-preserving synthetic data for recommendation systems. And it can provide privacy guarantees for the original private data while maximizing the utility of the synthetic data.

\subsection{Privacy Analysis (RQ2)}

\subsubsection{\textbf{Case Study}}
\begin{figure*}[ht]
	\centering
	{\includegraphics[width=0.9 \linewidth]{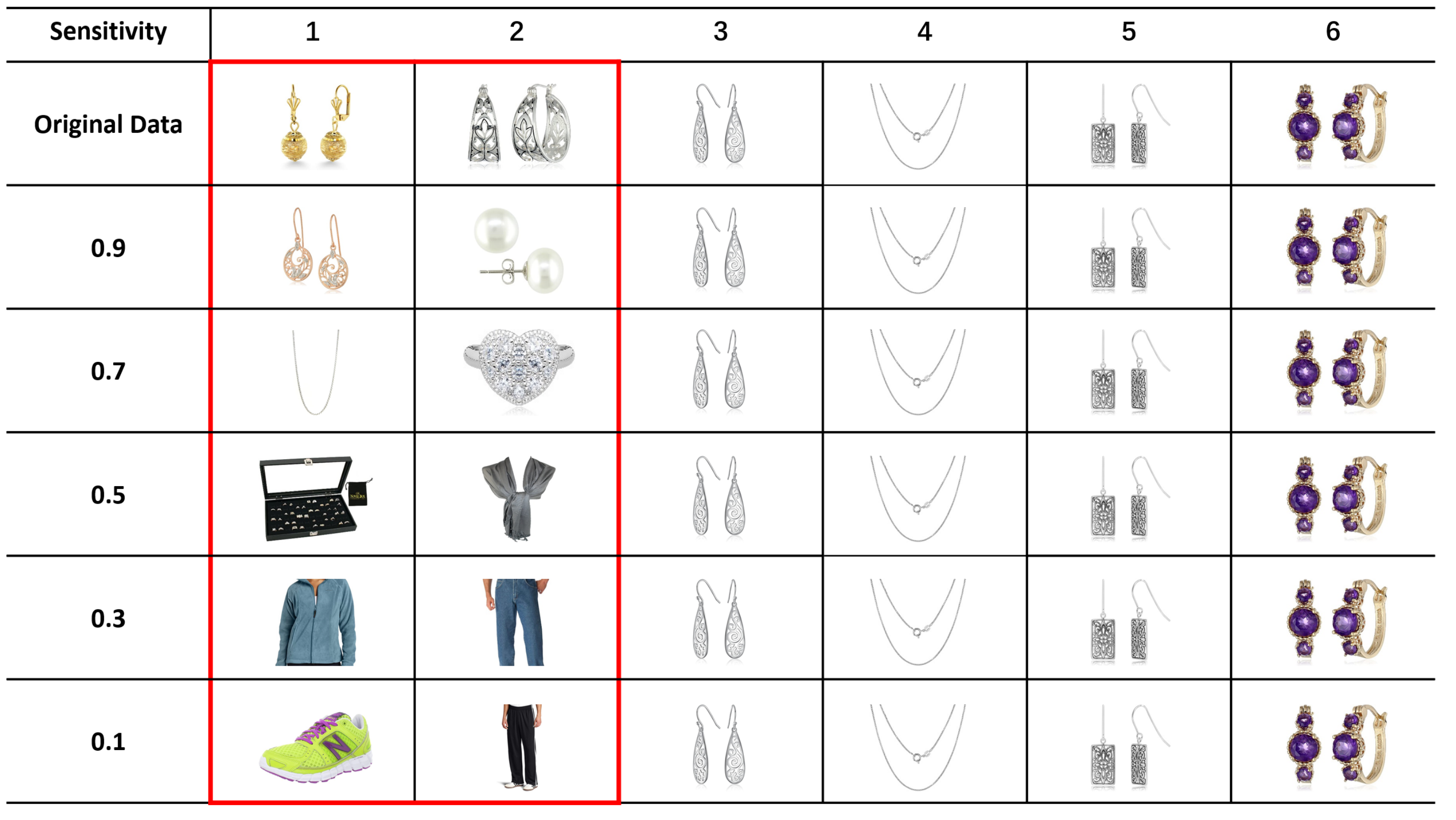}}
	\caption{Examples of privacy-preserving synthetic data on different sensitivities. The items that need to be replaced in the original and generated synthetic data under different sensitives are circled in the red box.}
	\vspace{0.0cm}
	\label{fig:visualization}
\end{figure*}
Employing the privacy-utility trade-off strategy, we can provide a privacy guarantee for the original item while maximizing the utility of the generated item. Towards this end, we selected one user and the user's interaction data both from \emph{Clothing} and the generated dataset based on \emph{Clothing} under specific privacy settings ($k = 0.2$). 

Figure~\ref{fig:visualization} shows the visualization of the original and generated data of the user.   
The original and generated user data are listed in rows. In contrast, the items in the original and generated user data are listed in columns. The items in the first two columns are the selected items in the original data and generated items in the generated user data. More specifically, the first row represents the user's original data from the Clothing dataset. From the perspective of the original items, we can find that the user likes jewelry a lot.

Our UPC-SDG model generates synthetic items by considering the user's preferences, the users' privacy preferences and the selected item's characteristics. From the items of the first two columns, it can be found that the generated synthetic items have more similar characteristics to the original item with increasing sensitivity. For example, when sensitivity $\gamma$ is set as $0.9$ and $0.7$, the generated items belong to the same category as that of the original items. 
With the decrease in sensitivity, the difference between the original and generated items becomes greater. Other than the differences between the original and generated items, we can also find that the generated item keeps up with the user's preferences. For instance, when $\gamma$ is set as $0.5$, the model is inclined to generate the accessory items in line with the user's preference. Finally, when $\gamma$ is set as $0.3$ and $0.1$, very different items are generated.

Overall, we perform visualization on the items to compare the original and generated user data, which demonstrates the effectiveness of our proposed model. In other words, the proposed model can remove the sensitive information from the user data by considering their privacy preference while maximizing the utility of the generated data. 

\begin{figure}[t]
	\centering
	\includegraphics[width=0.9\linewidth]{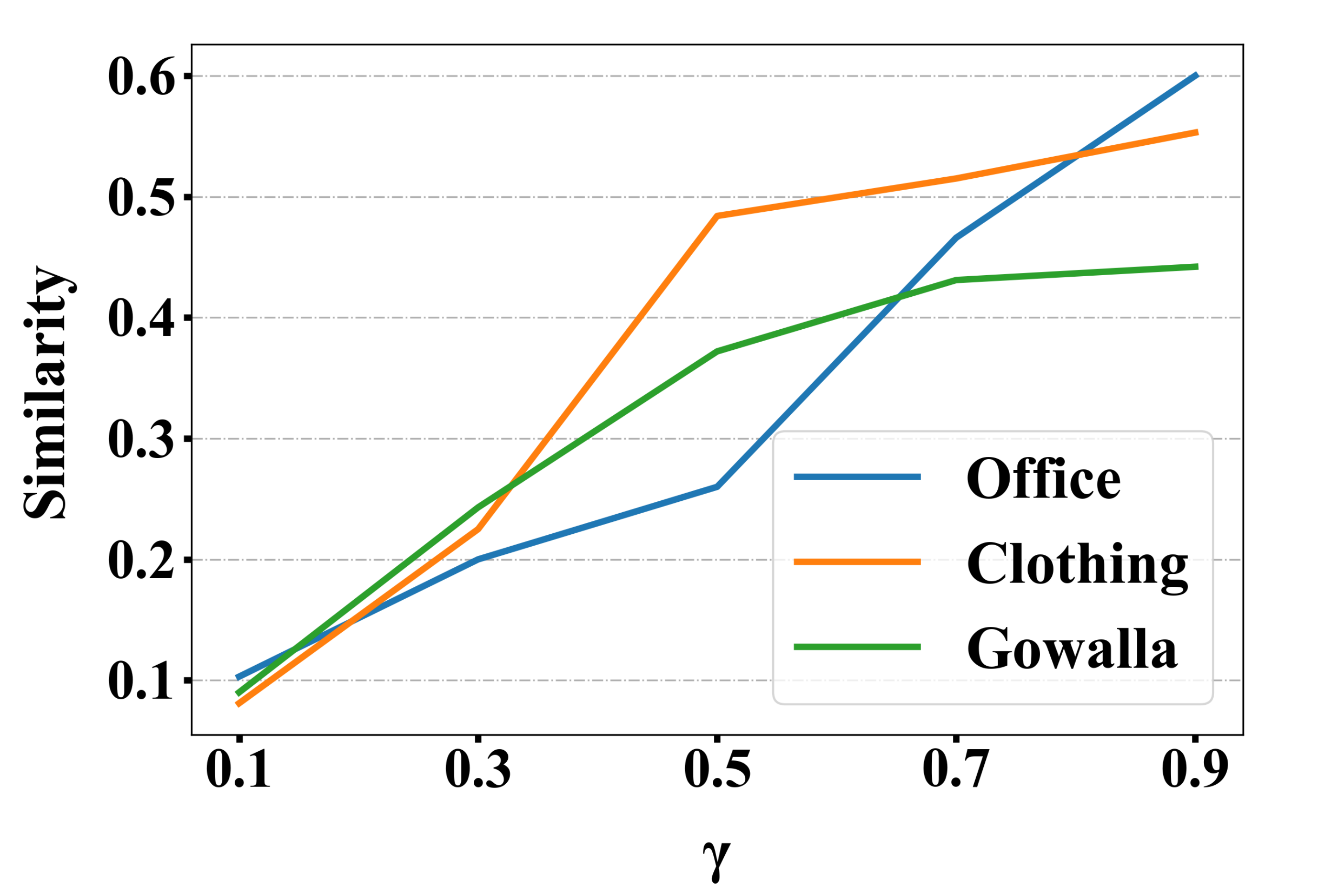}
	\caption{Visualization of the similarity between the generated synthetic items and original items. $\gamma$ denotes the sensitivity.}
	\label{fig:generated_similarity}
\end{figure}
\begin{table*}[ht]
	\caption{
		Performance of BPRMF over the datasets generated by our proposed model and its variants. The datasets are generated based on \emph{Office} under $\gamma = 0.9 $ and $ k_i \in \{0.2,0.4,0.6,0.8\}$, for $i=1,2,3,4$.
	}
	\begin{tabular}{l|ccc|ccc|ccc|ccc}
		\hline
		Dataset & \multicolumn{3}{c|}{Office$_{k1}$} & \multicolumn{3}{c|}{Office$_{k2}$} & \multicolumn{3}{c|}{Office$_{k3}$} & \multicolumn{3}{c}{Office$_{k4}$} \\ \hline
		Metrics     & PR        & Recall   & NDCG     & PR        & Recall   & NDCG     & PR        & Recall   & NDCG     & PR       & Recall   & NDCG     \\ \hline
		UPC-SDG$_s$ & 0.0028    & 0.0466   & 0.0215   & 0.0022    & 0.0356   & 0.0160   & 0.0016    & 0.0261   & 0.0118   & 0.0008   & 0.0143   & 0.0056   \\ \hline
		UPC-SDG$_r$ & 0.0030    & 0.0495   & 0.0227   & 0.0020    & 0.0325   & 0.0151   & 0.0014    & 0.0221   & 0.0098   & 0.0002   & 0.0027   & 0.0012   \\ \hline
		UPC-SDG$_f$ & 0.0030    & 0.0497   & 0.0229   & 0.0021    & 0.0338   & 0.0154   & 0.0016    & 0.0252   & 0.0109   & 0.0007   & 0.0105   & 0.0047   \\ \hline
		UPC-SDG & \textbf{0.0166}    & \textbf{0.0878}   & \textbf{0.0506}   & \textbf{0.0140}     & \textbf{0.0717}   & \textbf{0.0399}   & \textbf{0.0134}    & \textbf{0.0646}   & \textbf{0.0362}   & \textbf{0.0048}   & \textbf{0.0582}   & \textbf{0.0326}   \\ \hline
	\end{tabular}
	\label{tab:ablation}
	\vspace{0.0cm}
\end{table*}

\subsubsection{\textbf{Visualization of Similarity}}
For further analysis, we visualized the average similarity between the generated items and the original items for all users under different sensitivity settings.
As shown in Figure~\ref{fig:generated_similarity}, the x-axis represents the sensitivity that is preset by users before data generation, and the y-axis represents the average similarity calculated between the original items and the generated items. 
From the figure, the sensitivity and similarity are positively correlated over the three datasets. 
The results demonstrate that the proposed model can provide privacy guarantees for users' interaction data, and the privacy level is controllable.

\subsection{Ablation Study (RQ3)}

In this section, we examined the contributions of different components in our model to the final performance by comparing UPC-SDG with the following variants:
\begin{itemize}
\item \textbf{UPC-SDG$_s$} is a variant of our model which adopts a random strategy to replace the selection module.
\item \textbf{UPC-SDG$_r$} is a variant of our method which adopts a random strategy to replace the generation module.
\item \textbf{UPC-SDG$_f$} is a variant of our method which generates items by fixing the similarity between the selected item and the generated item.
\end{itemize}

The results of BPRMF over the datasets generated by the three variants and our model are reported in Table~\ref{tab:ablation}, in which the best results are highlighted in bold. We conducted ablation experiments over the generated dataset based on \emph{Office} under different replacement ratios and the same sensitivity\footnote{We omitted the results on \emph{Clothing} and \emph{Gowalla} because they show exactly the same trend.}. UPC-SDG$_f$ outperforms UPC-SDG$_r$ over all datasets, because the similarity between the generated item and the selected item is set as $0.9$. Due to the close similarity, recommendation models can learn better user representation by exploiting the generated items. It can be seen that our model consistently surpasses all variants across all metrics, which shows the reasonable design of our proposed model. 

\section{Conclusion}
In this paper, we present a User Privacy Controllable Synthetic Data Generation (UPC-SDG) model, which learns the characteristics of the original data and creates synthetic data based on users' privacy preferences for recommendation. Existing solutions are designed for addressing privacy issues in the training and results collection phase of recommendation systems. Users can still suffer from the problem of privacy leakage if their behavior data is either shared with organizations or are released publicly. Therefore,
the proposed UPC-SDG model aims to provide certain guarantees about privacy while maximizing the utility of synthetic data from both data level and item level perspectives. In light of this, at the data level, we design a selection module to select the items which contribute less to a user's preferences as the selected items need to be replaced.  At the item level, a synthetic data generation module is proposed to generate the synthetic items using the selected items by considering the user's preferences. Moreover, we also present a privacy-utility trade-off strategy to balance privacy and utility.
Extensive experiments have been conducted on three real-world datasets from Amazon and Gowalla. The experimental results demonstrate the effectiveness of our model. The ablation studies further validate the importance of the various components of our model.

\section{Acknowledgments}
This research is supported by the National Research Foundation, Singapore under its Strategic Capability Research Centres Funding Initiative, the National Natural Science Foundation of China, No.:61902223, No.:61906108 and Young creative team in universities of Shandong Province, No.:2020KJN012. Any opinions, findings and conclusions or recommendations expressed in this material are those of the author(s) and do not reflect the views of National Research Foundation, Singapore.

\bibliographystyle{ACM-Reference-Format}
\bibliography{privacy}


\begin{thebibliography}{49}


\ifx \showCODEN    \undefined \def \showCODEN     #1{\unskip}     \fi
\ifx \showDOI      \undefined \def \showDOI       #1{#1}\fi
\ifx \showISBNx    \undefined \def \showISBNx     #1{\unskip}     \fi
\ifx \showISBNxiii \undefined \def \showISBNxiii  #1{\unskip}     \fi
\ifx \showISSN     \undefined \def \showISSN      #1{\unskip}     \fi
\ifx \showLCCN     \undefined \def \showLCCN      #1{\unskip}     \fi
\ifx \shownote     \undefined \def \shownote      #1{#1}          \fi
\ifx \showarticletitle \undefined \def \showarticletitle #1{#1}   \fi
\ifx \showURL      \undefined \def \showURL       {\relax}        \fi
\providecommand\bibfield[2]{#2}
\providecommand\bibinfo[2]{#2}
\providecommand\natexlab[1]{#1}
\providecommand\showeprint[2][]{arXiv:#2}

\bibitem[Acs et~al\mbox{.}(2017)]%
        {Acs2017DPMGNN}
\bibfield{author}{\bibinfo{person}{Gergely Acs}, \bibinfo{person}{Luca Melis},
  \bibinfo{person}{Claude Castelluccia}, {and} \bibinfo{person}{Emiliano
  De~Cristofaro}.} \bibinfo{year}{2017}\natexlab{}.
\newblock \showarticletitle{Differentially Private Mixture of Generative Neural
  Networks}. In \bibinfo{booktitle}{\emph{ICDM}}. \bibinfo{pages}{715--720}.
\newblock


\bibitem[Barak et~al\mbox{.}(2007)]%
        {Barak2007PAC}
\bibfield{author}{\bibinfo{person}{Boaz Barak}, \bibinfo{person}{Kamalika
  Chaudhuri}, \bibinfo{person}{Cynthia Dwork}, \bibinfo{person}{Satyen Kale},
  \bibinfo{person}{Frank McSherry}, {and} \bibinfo{person}{Kunal Talwar}.}
  \bibinfo{year}{2007}\natexlab{}.
\newblock \showarticletitle{Privacy, Accuracy, and Consistency Too: A Holistic
  Solution to Contingency Table Release}. In \bibinfo{booktitle}{\emph{PODS}}.
  \bibinfo{publisher}{ACM}, \bibinfo{pages}{273–282}.
\newblock


\bibitem[Bindschaedler et~al\mbox{.}(2017)]%
        {Bindschaedler2017PDPPDS}
\bibfield{author}{\bibinfo{person}{Vincent Bindschaedler},
  \bibinfo{person}{Reza Shokri}, {and} \bibinfo{person}{Carl~A. Gunter}.}
  \bibinfo{year}{2017}\natexlab{}.
\newblock \showarticletitle{Plausible Deniability for Privacy-Preserving Data
  Synthesis}.
\newblock \bibinfo{journal}{\emph{Proc. VLDB Endow.}} (\bibinfo{year}{2017}),
  \bibinfo{pages}{481–492}.
\newblock


\bibitem[Cai et~al\mbox{.}(2018)]%
        {Cai2018GAN-HBNR}
\bibfield{author}{\bibinfo{person}{Xiaoyan Cai}, \bibinfo{person}{Junwei Han},
  {and} \bibinfo{person}{Libin Yang}.} \bibinfo{year}{2018}\natexlab{}.
\newblock \showarticletitle{Generative Adversarial Network Based Heterogeneous
  Bibliographic Network Representation for Personalized Citation
  Recommendation}. In \bibinfo{booktitle}{\emph{AAAI}}.
  \bibinfo{publisher}{AAAI Press}, \bibinfo{pages}{5747–5754}.
\newblock


\bibitem[Chae et~al\mbox{.}(2018)]%
        {Chae2018CFGANAG}
\bibfield{author}{\bibinfo{person}{Dong-Kyu Chae}, \bibinfo{person}{Jin-Soo
  Kang}, \bibinfo{person}{Sang-Wook Kim}, {and} \bibinfo{person}{Jung-Tae
  Lee}.} \bibinfo{year}{2018}\natexlab{}.
\newblock \showarticletitle{CFGAN: A Generic Collaborative Filtering Framework
  Based on Generative Adversarial Networks}. In
  \bibinfo{booktitle}{\emph{CIKM}}. \bibinfo{publisher}{ACM},
  \bibinfo{pages}{137–146}.
\newblock


\bibitem[Cheng et~al\mbox{.}(2018a)]%
        {Cheng2018A3ncf}
\bibfield{author}{\bibinfo{person}{Zhiyong Cheng}, \bibinfo{person}{Ying Ding},
  \bibinfo{person}{Xiangnan He}, \bibinfo{person}{Lei Zhu},
  \bibinfo{person}{Xuemeng Song}, {and} \bibinfo{person}{Mohan~S Kankanhalli}.}
  \bibinfo{year}{2018}\natexlab{a}.
\newblock \showarticletitle{A3NCF: An Adaptive Aspect Attention Model for
  Rating Prediction.}. In \bibinfo{booktitle}{\emph{IJCAI}}. Morgan Kaufmann,
  \bibinfo{pages}{3748--3754}.
\newblock


\bibitem[Cheng et~al\mbox{.}(2018b)]%
        {cheng2018aspect}
\bibfield{author}{\bibinfo{person}{Zhiyong Cheng}, \bibinfo{person}{Ying Ding},
  \bibinfo{person}{Lei Zhu}, {and} \bibinfo{person}{Kankanhalli Mohan}.}
  \bibinfo{year}{2018}\natexlab{b}.
\newblock \showarticletitle{Aspect-aware latent factor model: Rating prediction
  with ratings and reviews}. In \bibinfo{booktitle}{\emph{WWW}}. IW3C2,
  \bibinfo{pages}{639--648}.
\newblock


\bibitem[Cheng et~al\mbox{.}(2022)]%
        {Cheng2022FLA}
\bibfield{author}{\bibinfo{person}{Zhiyong Cheng}, \bibinfo{person}{Fan Liu},
  \bibinfo{person}{Shenghan Mei}, \bibinfo{person}{Yangyang Guo},
  \bibinfo{person}{Lei Zhu}, {and} \bibinfo{person}{Liqiang Nie}.}
  \bibinfo{year}{2022}\natexlab{}.
\newblock \showarticletitle{Feature-Level Attentive ICF for Recommendation}.
\newblock \bibinfo{journal}{\emph{TOIS}} \bibinfo{volume}{40},
  \bibinfo{number}{4}, \bibinfo{pages}{1--24}.
\newblock


\bibitem[Cunningham et~al\mbox{.}(2021)]%
        {Cunningham2021PPSLD}
\bibfield{author}{\bibinfo{person}{Teddy Cunningham}, \bibinfo{person}{Graham
  Cormode}, {and} \bibinfo{person}{Hakan Ferhatosmanoglu}.}
  \bibinfo{year}{2021}\natexlab{}.
\newblock \showarticletitle{Privacy-Preserving Synthetic Location Data in the
  Real World}. In \bibinfo{booktitle}{\emph{SSTD}}. \bibinfo{publisher}{ACM},
  \bibinfo{pages}{23–33}.
\newblock


\bibitem[Fan and Kankanhalli(2021)]%
        {Hehefan2021Motion}
\bibfield{author}{\bibinfo{person}{Hehe Fan} {and} \bibinfo{person}{Mohan
  Kankanhalli}.} \bibinfo{year}{2021}\natexlab{}.
\newblock \showarticletitle{Motion = Video - Content: Towards Unsupervised
  Learning of Motion Representation from Videos}. In
  \bibinfo{booktitle}{\emph{ACM Multimedia Asia}}. \bibinfo{publisher}{ACM},
  \bibinfo{pages}{1--7}.
\newblock


\bibitem[Fan et~al\mbox{.}(2019)]%
        {Fan2019DASO}
\bibfield{author}{\bibinfo{person}{Wenqi Fan}, \bibinfo{person}{Tyler Derr},
  \bibinfo{person}{Yao Ma}, \bibinfo{person}{Jianping Wang},
  \bibinfo{person}{Jiliang Tang}, {and} \bibinfo{person}{Qing Li}.}
  \bibinfo{year}{2019}\natexlab{}.
\newblock \showarticletitle{Deep Adversarial Social Recommendation}. In
  \bibinfo{booktitle}{\emph{IJCAI}}. \bibinfo{publisher}{AAAI Press},
  \bibinfo{pages}{1351–1357}.
\newblock


\bibitem[Gao et~al\mbox{.}(2020)]%
        {Gao2020DPLCF}
\bibfield{author}{\bibinfo{person}{Chen Gao}, \bibinfo{person}{Chao Huang},
  \bibinfo{person}{Dongsheng Lin}, \bibinfo{person}{Depeng Jin}, {and}
  \bibinfo{person}{Yong Li}.} \bibinfo{year}{2020}\natexlab{}.
\newblock \showarticletitle{DPLCF: Differentially Private Local Collaborative
  Filtering}. In \bibinfo{booktitle}{\emph{SIGIR}}. \bibinfo{publisher}{ACM},
  \bibinfo{pages}{961–970}.
\newblock


\bibitem[Gao et~al\mbox{.}(2021)]%
        {RS2021problem}
\bibfield{author}{\bibinfo{person}{Min Gao}, \bibinfo{person}{Junwei Zhang},
  \bibinfo{person}{Junliang Yu}, \bibinfo{person}{Jundong Li},
  \bibinfo{person}{Junhao Wen}, {and} \bibinfo{person}{Qingyu Xiong}.}
  \bibinfo{year}{2021}\natexlab{}.
\newblock \showarticletitle{Recommender systems based on generative adversarial
  networks: A problem-driven perspective}.
\newblock \bibinfo{journal}{\emph{INS}}  \bibinfo{volume}{546}
  (\bibinfo{year}{2021}), \bibinfo{pages}{1166--1185}.
\newblock


\bibitem[Guo et~al\mbox{.}(2020)]%
        {tkde-fm}
\bibfield{author}{\bibinfo{person}{Yangyang Guo}, \bibinfo{person}{Zhiyong
  Cheng}, \bibinfo{person}{Jiazheng Jing}, \bibinfo{person}{Yanpeng Lin},
  \bibinfo{person}{Liqiang Nie}, {and} \bibinfo{person}{Meng Wang}.}
  \bibinfo{year}{2020}\natexlab{}.
\newblock \showarticletitle{Enhancing Factorization Machines with Generalized
  Metric Learning}.
\newblock \bibinfo{journal}{\emph{TKDE}} (\bibinfo{year}{2020}),
  \bibinfo{pages}{1--1}.
\newblock


\bibitem[Havrylov and Titov(2017)]%
        {Gumbelsoftmax2017nips}
\bibfield{author}{\bibinfo{person}{Serhii Havrylov} {and} \bibinfo{person}{Ivan
  Titov}.} \bibinfo{year}{2017}\natexlab{}.
\newblock \showarticletitle{Emergence of Language with Multi-Agent Games:
  Learning to Communicate with Sequences of Symbols}. In
  \bibinfo{booktitle}{\emph{NIPS}}. \bibinfo{publisher}{CAI.},
  \bibinfo{pages}{2146–2156}.
\newblock


\bibitem[He et~al\mbox{.}(2015)]%
        {he2015trirank}
\bibfield{author}{\bibinfo{person}{Xiangnan He}, \bibinfo{person}{Tao Chen},
  \bibinfo{person}{Min-Yen Kan}, {and} \bibinfo{person}{Xiao Chen}.}
  \bibinfo{year}{2015}\natexlab{}.
\newblock \showarticletitle{TriRank: Review aware Explainable Recommendation by
  Modeling Aspects}. In \bibinfo{booktitle}{\emph{CIKM}}.
  \bibinfo{publisher}{ACM}, \bibinfo{pages}{1661--1670}.
\newblock


\bibitem[He et~al\mbox{.}(2020)]%
        {He2020lightgcn}
\bibfield{author}{\bibinfo{person}{Xiangnan He}, \bibinfo{person}{Kuan Deng},
  \bibinfo{person}{Xiang Wang}, \bibinfo{person}{Yan Li},
  \bibinfo{person}{Yong{-}Dong Zhang}, {and} \bibinfo{person}{Meng Wang}.}
  \bibinfo{year}{2020}\natexlab{}.
\newblock \showarticletitle{LightGCN: Simplifying and Powering Graph
  Convolution Network for Recommendation}. In
  \bibinfo{booktitle}{\emph{SIGIR}}. ACM, \bibinfo{pages}{639--648}.
\newblock


\bibitem[He et~al\mbox{.}(2018a)]%
        {He2018Adversarial}
\bibfield{author}{\bibinfo{person}{Xiangnan He}, \bibinfo{person}{Zhankui He},
  \bibinfo{person}{Xiaoyu Du}, {and} \bibinfo{person}{Tat-Seng Chua}.}
  \bibinfo{year}{2018}\natexlab{a}.
\newblock \showarticletitle{Adversarial personalized ranking for
  recommendation}. In \bibinfo{booktitle}{\emph{SIGIR}}. ACM,
  \bibinfo{pages}{355--364}.
\newblock


\bibitem[He et~al\mbox{.}(2018b)]%
        {He2018Nais}
\bibfield{author}{\bibinfo{person}{Xiangnan He}, \bibinfo{person}{Zhankui He},
  \bibinfo{person}{Jingkuan Song}, \bibinfo{person}{Zhenguang Liu},
  \bibinfo{person}{Yu-Gang Jiang}, {and} \bibinfo{person}{Tat-Seng Chua}.}
  \bibinfo{year}{2018}\natexlab{b}.
\newblock \showarticletitle{NAIS: Neural attentive item similarity model for
  recommendation}.
\newblock \bibinfo{journal}{\emph{TKDE}}  \bibinfo{volume}{30}
  (\bibinfo{year}{2018}), \bibinfo{pages}{2354--2366}.
\newblock


\bibitem[He et~al\mbox{.}(2017)]%
        {He2017Neural}
\bibfield{author}{\bibinfo{person}{Xiangnan He}, \bibinfo{person}{Lizi Liao},
  \bibinfo{person}{Hanwang Zhang}, \bibinfo{person}{Liqiang Nie},
  \bibinfo{person}{Xia Hu}, {and} \bibinfo{person}{Tat-Seng Chua}.}
  \bibinfo{year}{2017}\natexlab{}.
\newblock \showarticletitle{Neural collaborative filtering}. In
  \bibinfo{booktitle}{\emph{WWW}}. ACM, \bibinfo{pages}{173--182}.
\newblock


\bibitem[Hsieh et~al\mbox{.}(2017)]%
        {hsieh2017collaborative}
\bibfield{author}{\bibinfo{person}{Cheng-Kang Hsieh}, \bibinfo{person}{Longqi
  Yang}, \bibinfo{person}{Yin Cui}, \bibinfo{person}{Tsung-Yi Lin},
  \bibinfo{person}{Serge Belongie}, {and} \bibinfo{person}{Deborah Estrin}.}
  \bibinfo{year}{2017}\natexlab{}.
\newblock \showarticletitle{Collaborative metric learning}. In
  \bibinfo{booktitle}{\emph{WWW}}. IW3C2, \bibinfo{pages}{193--201}.
\newblock


\bibitem[Hu et~al\mbox{.}(2008)]%
        {Hu2008Collaborative}
\bibfield{author}{\bibinfo{person}{Yifan Hu}, \bibinfo{person}{Yehuda Koren},
  {and} \bibinfo{person}{Chris Volinsky}.} \bibinfo{year}{2008}\natexlab{}.
\newblock \showarticletitle{Collaborative filtering for implicit feedback
  datasets}. In \bibinfo{booktitle}{\emph{ICDM}}. IEEE,
  \bibinfo{pages}{263--272}.
\newblock


\bibitem[Jang et~al\mbox{.}(2017)]%
        {jang2017categorical}
\bibfield{author}{\bibinfo{person}{Eric Jang}, \bibinfo{person}{Shixiang Gu},
  {and} \bibinfo{person}{Ben Poole}.} \bibinfo{year}{2017}\natexlab{}.
\newblock \showarticletitle{Categorical Reparameterization with
  Gumbel-Softmax}.
\newblock  (\bibinfo{year}{2017}).
\newblock


\bibitem[Kingma and Ba(2015)]%
        {kingma2014adam}
\bibfield{author}{\bibinfo{person}{Diederik~P Kingma} {and}
  \bibinfo{person}{Jimmy Ba}.} \bibinfo{year}{2015}\natexlab{}.
\newblock \showarticletitle{Adam: A method for stochastic optimization}. In
  \bibinfo{booktitle}{\emph{ICLR}}.
\newblock


\bibitem[Koren(2008)]%
        {svd++}
\bibfield{author}{\bibinfo{person}{Yehuda Koren}.}
  \bibinfo{year}{2008}\natexlab{}.
\newblock \showarticletitle{Factorization meets the neighborhood: a
  multifaceted collaborative filtering model}. In
  \bibinfo{booktitle}{\emph{SIGKDD}}. \bibinfo{publisher}{ACM},
  \bibinfo{pages}{426--434}.
\newblock


\bibitem[Koren(2010)]%
        {Koren2010Factor}
\bibfield{author}{\bibinfo{person}{Yehuda Koren}.}
  \bibinfo{year}{2010}\natexlab{}.
\newblock \showarticletitle{Factor in the neighbors: Scalable and accurate
  collaborative filtering}.
\newblock \bibinfo{journal}{\emph{TKDD}}  \bibinfo{volume}{4}
  (\bibinfo{year}{2010}), \bibinfo{pages}{1--24}.
\newblock


\bibitem[Liu et~al\mbox{.}(2019)]%
        {liu2019mm}
\bibfield{author}{\bibinfo{person}{Fan Liu}, \bibinfo{person}{Zhiyong Cheng},
  \bibinfo{person}{Changchang Sun}, \bibinfo{person}{Yinglong Wang},
  \bibinfo{person}{Liqiang Nie}, {and} \bibinfo{person}{Mohan~S. Kankanhalli}.}
  \bibinfo{year}{2019}\natexlab{}.
\newblock \showarticletitle{User Diverse Preference Modeling by Multimodal
  Attentive Metric Learning}. In \bibinfo{booktitle}{\emph{MM}}.
  \bibinfo{publisher}{{ACM}}, \bibinfo{pages}{1526--1534}.
\newblock


\bibitem[Liu et~al\mbox{.}(2021)]%
        {liu2021interest}
\bibfield{author}{\bibinfo{person}{Fan Liu}, \bibinfo{person}{Zhiyong Cheng},
  \bibinfo{person}{Lei Zhu}, \bibinfo{person}{Zan Gao}, {and}
  \bibinfo{person}{Liqiang Nie}.} \bibinfo{year}{2021}\natexlab{}.
\newblock \showarticletitle{Interest-aware Message-Passing {GCN} for
  Recommendation}. In \bibinfo{booktitle}{\emph{WWW}}.
  \bibinfo{publisher}{IW3C2}, \bibinfo{pages}{1296--1305}.
\newblock


\bibitem[Liu et~al\mbox{.}(2020)]%
        {Liu2020A2GCN}
\bibfield{author}{\bibinfo{person}{Fan Liu}, \bibinfo{person}{Zhiyong Cheng},
  \bibinfo{person}{Lei Zhu}, \bibinfo{person}{Chenghao Liu}, {and}
  \bibinfo{person}{Liqiang Nie}.} \bibinfo{year}{2020}\natexlab{}.
\newblock \showarticletitle{A2-GCN: An Attribute-aware Attentive GCN Model for
  Recommendation}. In \bibinfo{booktitle}{\emph{TKDE}}. \bibinfo{pages}{1--1}.
\newblock


\bibitem[Liu et~al\mbox{.}(2022)]%
        {liu2022review}
\bibfield{author}{\bibinfo{person}{Han Liu}, \bibinfo{person}{Yangyang Guo},
  \bibinfo{person}{Jianhua Yin}, \bibinfo{person}{Zan Gao}, {and}
  \bibinfo{person}{Liqiang Nie}.} \bibinfo{year}{2022}\natexlab{}.
\newblock \showarticletitle{Review polarity-wise recommender}.
\newblock \bibinfo{journal}{\emph{IEEE TNNLS}} (\bibinfo{year}{2022}),
  \bibinfo{pages}{1--1}.
\newblock


\bibitem[Pan et~al\mbox{.}(2008)]%
        {Pan2008One}
\bibfield{author}{\bibinfo{person}{Rong Pan}, \bibinfo{person}{Yunhong Zhou},
  \bibinfo{person}{Bin Cao}, \bibinfo{person}{Nathan~N Liu},
  \bibinfo{person}{Rajan Lukose}, \bibinfo{person}{Martin Scholz}, {and}
  \bibinfo{person}{Qiang Yang}.} \bibinfo{year}{2008}\natexlab{}.
\newblock \showarticletitle{One-class collaborative filtering}. In
  \bibinfo{booktitle}{\emph{ICDM}}. IEEE, \bibinfo{pages}{502--511}.
\newblock


\bibitem[Qiang(2019)]%
        {Yang2019FRS}
\bibfield{author}{\bibinfo{person}{Yang Qiang}.}
  \bibinfo{year}{2019}\natexlab{}.
\newblock \showarticletitle{Federated Recommendation Systems}. In
  \bibinfo{booktitle}{\emph{ICBD}}. \bibinfo{pages}{1--1}.
\newblock


\bibitem[Rafailidis and Crestani(2019)]%
        {Rafailidis2019ATR}
\bibfield{author}{\bibinfo{person}{Dimitrios Rafailidis} {and}
  \bibinfo{person}{Fabio Crestani}.} \bibinfo{year}{2019}\natexlab{}.
\newblock \showarticletitle{Adversarial Training for Review-Based
  Recommendations}. In \bibinfo{booktitle}{\emph{SIGIR}}.
  \bibinfo{publisher}{ACM}, \bibinfo{pages}{1057–1060}.
\newblock


\bibitem[Rendle et~al\mbox{.}(2009)]%
        {Rendle2009Bpr}
\bibfield{author}{\bibinfo{person}{Steffen Rendle}, \bibinfo{person}{Christoph
  Freudenthaler}, \bibinfo{person}{Zeno Gantner}, {and} \bibinfo{person}{Lars
  Schmidt-Thieme}.} \bibinfo{year}{2009}\natexlab{}.
\newblock \showarticletitle{BPR: Bayesian personalized ranking from implicit
  feedback}. In \bibinfo{booktitle}{\emph{UAI}}. AUAI,
  \bibinfo{pages}{452--461}.
\newblock


\bibitem[Salakhutdinov and Mnih(2007)]%
        {Salak2007PMF}
\bibfield{author}{\bibinfo{person}{Ruslan Salakhutdinov} {and}
  \bibinfo{person}{Andriy Mnih}.} \bibinfo{year}{2007}\natexlab{}.
\newblock \showarticletitle{Probabilistic Matrix Factorization}. In
  \bibinfo{booktitle}{\emph{NIPS}}. \bibinfo{publisher}{CAI.},
  \bibinfo{pages}{1257–1264}.
\newblock


\bibitem[Sweeney(2002)]%
        {Sweeney2002MPP}
\bibfield{author}{\bibinfo{person}{Latanya Sweeney}.}
  \bibinfo{year}{2002}\natexlab{}.
\newblock \showarticletitle{K-Anonymity: A Model for Protecting Privacy}.
\newblock \bibinfo{journal}{\emph{UFKS}} \bibinfo{volume}{10},
  \bibinfo{number}{5} (\bibinfo{year}{2002}), \bibinfo{pages}{557–570}.
\newblock


\bibitem[Tong et~al\mbox{.}(2019)]%
        {Tong2019CollaborativeGA}
\bibfield{author}{\bibinfo{person}{Yuzhen Tong}, \bibinfo{person}{Yadan Luo},
  \bibinfo{person}{Zheng Zhang}, \bibinfo{person}{Shazia Sadiq}, {and}
  \bibinfo{person}{Peng Cui}.} \bibinfo{year}{2019}\natexlab{}.
\newblock \showarticletitle{Collaborative Generative Adversarial Network for
  Recommendation Systems}. In \bibinfo{booktitle}{\emph{ICDE Workshops}}.
  \bibinfo{pages}{161--168}.
\newblock


\bibitem[Tran et~al\mbox{.}(2019)]%
        {Tran2019MASR}
\bibfield{author}{\bibinfo{person}{Thanh Tran}, \bibinfo{person}{Renee
  Sweeney}, {and} \bibinfo{person}{Kyumin Lee}.}
  \bibinfo{year}{2019}\natexlab{}.
\newblock \showarticletitle{Adversarial Mahalanobis Distance-Based Attentive
  Song Recommender for Automatic Playlist Continuation}. In
  \bibinfo{booktitle}{\emph{SIGIR}}. \bibinfo{publisher}{ACM},
  \bibinfo{pages}{245–254}.
\newblock


\bibitem[van~den Berg et~al\mbox{.}(2018)]%
        {berg2019gcmc}
\bibfield{author}{\bibinfo{person}{Rianne van~den Berg},
  \bibinfo{person}{Thomas~N. Kipf}, {and} \bibinfo{person}{Max Welling}.}
  \bibinfo{year}{2018}\natexlab{}.
\newblock \showarticletitle{Graph Convolutional Matrix Completion}. In
  \bibinfo{booktitle}{\emph{SIGKDD: Deep Learning Day}}.
  \bibinfo{publisher}{ACM}.
\newblock


\bibitem[Wang and Tang(2017)]%
        {Wang2017DPRS}
\bibfield{author}{\bibinfo{person}{Jun Wang} {and} \bibinfo{person}{Qiang
  Tang}.} \bibinfo{year}{2017}\natexlab{}.
\newblock \showarticletitle{Differentially Private Neighborhood-Based
  Recommender Systems}. In \bibinfo{booktitle}{\emph{SEC}}.
  \bibinfo{publisher}{SIP}, \bibinfo{pages}{459--473}.
\newblock


\bibitem[Wang et~al\mbox{.}(2018)]%
        {Wang2018NMRN-GAN}
\bibfield{author}{\bibinfo{person}{Qinyong Wang}, \bibinfo{person}{Hongzhi
  Yin}, \bibinfo{person}{Zhiting Hu}, \bibinfo{person}{Defu Lian},
  \bibinfo{person}{Hao Wang}, {and} \bibinfo{person}{Zi Huang}.}
  \bibinfo{year}{2018}\natexlab{}.
\newblock \showarticletitle{Neural Memory Streaming Recommender Networks with
  Adversarial Training}. In \bibinfo{booktitle}{\emph{SIGKDD}}.
  \bibinfo{publisher}{ACM}, \bibinfo{pages}{2467–2475}.
\newblock


\bibitem[Wang et~al\mbox{.}(2019b)]%
        {Wang2019EnhancingCF}
\bibfield{author}{\bibinfo{person}{Qinyong Wang}, \bibinfo{person}{Hongzhi
  Yin}, \bibinfo{person}{Hao Wang}, \bibinfo{person}{Quoc Viet~Hung Nguyen},
  \bibinfo{person}{Zi Huang}, {and} \bibinfo{person}{Lizhen Cui}.}
  \bibinfo{year}{2019}\natexlab{b}.
\newblock \showarticletitle{Enhancing Collaborative Filtering with Generative
  Augmentation}. In \bibinfo{booktitle}{\emph{SIGKDD}}.
  \bibinfo{publisher}{ACM}, \bibinfo{pages}{548–556}.
\newblock


\bibitem[Wang et~al\mbox{.}(2019a)]%
        {wang2019ngcf}
\bibfield{author}{\bibinfo{person}{Xiang Wang}, \bibinfo{person}{Xiangnan He},
  \bibinfo{person}{Meng Wang}, \bibinfo{person}{Fuli Feng}, {and}
  \bibinfo{person}{Tat-Seng Chua}.} \bibinfo{year}{2019}\natexlab{a}.
\newblock \showarticletitle{Neural graph collaborative filtering}. In
  \bibinfo{booktitle}{\emph{SIGIR}}. \bibinfo{publisher}{ACM},
  \bibinfo{pages}{165--174}.
\newblock


\bibitem[Wei et~al\mbox{.}(2019)]%
        {wei2019hashtag}
\bibfield{author}{\bibinfo{person}{Yinwei Wei}, \bibinfo{person}{Zhiyong
  Cheng}, \bibinfo{person}{Xuzheng Yu}, \bibinfo{person}{Zhou Zhao},
  \bibinfo{person}{Lei Zhu}, {and} \bibinfo{person}{Liqiang Nie}.}
  \bibinfo{year}{2019}\natexlab{}.
\newblock \showarticletitle{Personalized Hashtag Recommendation for
  Micro-videos}. In \bibinfo{booktitle}{\emph{MM}}. \bibinfo{publisher}{{ACM}},
  \bibinfo{pages}{1446--1454}.
\newblock


\bibitem[Wei et~al\mbox{.}(2020)]%
        {Wei2019GRCN}
\bibfield{author}{\bibinfo{person}{Yinwei Wei}, \bibinfo{person}{Xiang Wang},
  \bibinfo{person}{Liqiang Nie}, \bibinfo{person}{Xiangnan He}, {and}
  \bibinfo{person}{Tat-Seng Chua}.} \bibinfo{year}{2020}\natexlab{}.
\newblock \showarticletitle{Graph-Refined Convolutional Network for Multimedia
  Recommendation with Implicit Feedback}. In \bibinfo{booktitle}{\emph{MM}}.
  \bibinfo{publisher}{ACM}, \bibinfo{pages}{3541–3549}.
\newblock


\bibitem[Xue et~al\mbox{.}(2017)]%
        {xue2017deep}
\bibfield{author}{\bibinfo{person}{Hong-Jian Xue}, \bibinfo{person}{Xinyu Dai},
  \bibinfo{person}{Jianbing Zhang}, \bibinfo{person}{Shujian Huang}, {and}
  \bibinfo{person}{Jiajun Chen}.} \bibinfo{year}{2017}\natexlab{}.
\newblock \showarticletitle{Deep Matrix Factorization Models for Recommender
  Systems.}. In \bibinfo{booktitle}{\emph{IJCAI}}. \bibinfo{publisher}{AAAI
  Press}, \bibinfo{pages}{3203--3209}.
\newblock


\bibitem[Yuan et~al\mbox{.}(2019)]%
        {Yuan2019ACAE}
\bibfield{author}{\bibinfo{person}{Feng Yuan}, \bibinfo{person}{Lina Yao},
  {and} \bibinfo{person}{Boualem Benatallah}.} \bibinfo{year}{2019}\natexlab{}.
\newblock \showarticletitle{Adversarial Collaborative Auto-encoder for Top-N
  Recommendation}. In \bibinfo{booktitle}{\emph{IJCNN}}. \bibinfo{pages}{1--8}.
\newblock


\bibitem[Zhao et~al\mbox{.}(2018)]%
        {Wei2018PLASTIC}
\bibfield{author}{\bibinfo{person}{Wei Zhao}, \bibinfo{person}{Benyou Wang},
  \bibinfo{person}{Jianbo Ye}, \bibinfo{person}{Yongqiang Gao},
  \bibinfo{person}{Min Yang}, {and} \bibinfo{person}{Xiaojun Chen}.}
  \bibinfo{year}{2018}\natexlab{}.
\newblock \showarticletitle{PLASTIC: Prioritize Long and Short-term Information
  in Top-n Recommendation using Adversarial Training}. In
  \bibinfo{booktitle}{\emph{IJCAI}}. \bibinfo{publisher}{AAAI Press},
  \bibinfo{pages}{3676--3682}.
\newblock


\bibitem[Zheng et~al\mbox{.}(2019)]%
        {Zhangnan2019tryon}
\bibfield{author}{\bibinfo{person}{Na Zheng}, \bibinfo{person}{Xuemeng Song},
  \bibinfo{person}{Zhaozheng Chen}, \bibinfo{person}{Linmei Hu},
  \bibinfo{person}{Da Cao}, {and} \bibinfo{person}{Liqiang Nie}.}
  \bibinfo{year}{2019}\natexlab{}.
\newblock \showarticletitle{Virtually Trying on New Clothing with Arbitrary
  Poses}. In \bibinfo{booktitle}{\emph{MM}}. \bibinfo{publisher}{ACM},
  \bibinfo{pages}{266–274}.
\newblock


\end{thebibliography}

\end{document}